\documentclass[fleqn,usenatbib]{mnras}

\usepackage{newtxtext,newtxmath}

\usepackage[T1]{fontenc}

\DeclareRobustCommand{\VAN}[3]{#2}
\let\VANthebibliography\thebibliography
\def\thebibliography{\DeclareRobustCommand{\VAN}[3]{##3}\VANthebibliography}

\usepackage{graphicx}	
\usepackage{amsmath}
\usepackage{soul} 

\title[Cyg~X-1, MAXI~J1820+070 spectral-timing]{Long term variability of Cygnus X-1. IX. A spectral-timing comparison of Cygnus X-1 and MAXI J1820+070 in the hard state}

\author[A. Basak et al.]
{
Arkadip Basak$^{1}$, Phil Uttley$^{1}$\thanks{E-mail: p.uttley@uva.nl}, Niek Bollemeijer$^{1}$, Matteo Bachetti$^{2}$, Arash Bahramian$^{3}$, Victoria Grinberg$^{4}$, \and
Erin Kara$^{5}$, Eleonora V. Lai$^{2}$, Thomas J. Maccarone$^{6}$, Barbara De Marco$^{7}$, James Miller-Jones$^{3}$, \and
Katja Pottschmidt $^{8,9}$\thanks{Deceased}, Simon A. Vaughan$^{10}$ and J{\"o}rn Wilms $^{11}$ \and 
\\
$^{1}$ Anton Pannekoek Institute for Astronomy, University of Amsterdam, 1098 XH, Amsterdam, The Netherlands \\
$^{2}$ INAF-Osservatorio Astronomico di Cagliari, Via della Scienza 5, I-09047, Selargius, CA, Italy \\
$^{3}$ International Centre for Radio Astronomy Research – Curtin University, Perth, WA 6845, Australia \\
$^{4}$ European Space Agency (ESA), European Space Research and Technology Centre (ESTEC), Keplerlaan 1, 2201 AZ, Noordwijk, The Netherlands \\
$^{5}$ MIT Kavli Institute for Astrophysics and Space Research, MIT, 77 Massachusetts Avenue, Cambridge, MA 02139, USA \\
$^{6}$ Department of Physics \& Astronomy, Texas Tech University, Box 41051, Lubbock, TX, 79409-1051, USA \\
$^{7}$ Departament de Fisíca, EEBE, Universitat Politècnica de Catalunya, Av. Eduard Maristany 16, 08019 Barcelona, Spain \\
$^{8}$ Center for Space Sciences and Technology, University of Maryland,
Baltimore County, 1000 Hilltop Circle, Baltimore, MD 21250, USA \\
$^{9}$ CRESST and NASA Goddard Space Flight Center, Astroparticle Physics
Laboratory, 8800 Greenbelt Road, Greenbelt, MD 20771, USA \\
$^{10}$ School of Physics and Astronomy, University of Leicester, University Road, Leicester, LE1 7RH, UK \\
$^{11}$ Dr. Karl Remeis-Observatory, University of Erlangen-Nuremberg, Sternwartstr. 7, 96049, Bamberg, Germany \\
}

\date{Accepted XXX. Received YYY; in original form ZZZ}

\pubyear{2024}

\begin{document}
\label{firstpage}
\pagerange{\pageref{firstpage}--\pageref{lastpage}}
\maketitle

\begin{abstract}
Cygnus X-1 is a persistent, high-mass black hole X-ray binary (BHXRB) which in the hard state shows many similar properties to transient BHXRBs, along with intriguing differences, such as the lack of quasi-periodic oscillations. Here, we compare for the first time the detailed spectral-timing properties of Cyg~X-1 with a transient BHXRB, MAXI~J1820+070, combining data from {\it XMM-Newton} and {\it NICER} with contemporaneous {\it INTEGRAL} data to study the power spectra, rms spectra and time-lags over a broad 0.5--200~keV range. We select bright hard state MAXI~J1820+070 data with similar power-spectral shapes to the Cyg~X-1 data, to compare the source behaviours while accounting for the evolution of spectral-timing properties, notably the lags, through the hard state. Cyg~X-1 shows no evidence for soft lags in the 1--10~Hz frequency range where they are clearly detected for MAXI~J1820+070. Furthermore the low-frequency hard lags and rms-spectra evolve much more strongly during the hard state of Cyg~X-1 than for MAXI~J1820+070. We argue that these differences cannot be explained by the different black hole masses of these systems, but may be related to their different accretion rates and corresponding locations on the hardness-intensity diagram. We conjecture that there is a significant luminosity-dependence of coronal geometry in the hard state of BHXRBs, rather than an intrinsic difference between Cyg~X-1 and transient BHXRBs. This possibility has also been suggested to explain a common time-lag feature that appears in the hard intermediate states of Cyg~X-1 and transient BHXRBs.
\end{abstract}

\begin{keywords}
accretion, accretion discs -- black hole physics -- X-rays: binaries
\end{keywords}



\section{Introduction}
\label{section:Intro}

Cygnus~X-1 (Cyg~X-1) is the most massive confirmed black hole X-ray binary (BHXRB), with a black hole of mass $M_{\rm BH}\simeq$ 21.2 $\pm$ 2.2 $M_{\odot}$ \citep{jones21}, accreting matter from the wind of an O9.7 Iab companion star (HDE 226868). As a persistent X-ray source and the earliest-known BHXRB \citep{WebsterMurdin1972}, Cyg~X-1 has often served as a template for BHXRB X-ray properties, even though it shows a number of characteristics which are distinct when compared to the more numerous transient BHXRBs, which have low mass stellar companions. For example, separate hard and soft X-ray spectral states correlated with radio jet emission were first identified in Cyg X-1 \citep{Tananbaumetal1972} and later found to be common to all transient BHXRBs in outburst (e.g. see the review by \citealt{FenderMunozDarias2016}). The hard state spectrum appears similar to that of other BHXRBs, dominated primarily by a hard power-law like component with photon index $\Gamma\simeq 1.5$--$2$ and most likely originating in a Comptonizing hot corona \citep{Kalemcietal2022}. The corona upscatters seed photons from an accretion disc which produces a relatively weaker blackbody component that can be observed in soft X-rays, corresponding to an inner disc temperature $<0.3$~keV. As for transient BHXRBs, the soft state of Cyg X-1 shows a steeper power-law ($\Gamma > 2$) and stronger disc blackbody emission than the hard state \citep{Churazovetal2001}. However the Cyg~X-1 soft state shows a lower disc temperature and relatively stronger power-law emission, when compared with the typical X-ray spectra of transient BHXRB soft states \citep{Kawanoetal2017}.

Timing behaviour also shows both similarities and differences between Cyg~X-1 and transient BHXRBs. The hard state power spectra of Cyg~X-1 and transient BHXRBs show characteristic broadband noise that can be reasonably well described by a sum of 3 or more broad Lorentzians \citep{potts03,KleinWoltvdKlis2008}. As Cyg~X-1 and transient BHXRBs evolve through the hard state and the intermediate spectral states \citep{Bellonietal2005}, the characteristic low-frequency break increases in frequency, and the power spectrum becomes more band-limited as a result \citep{potts03,Heil15}. However, transient BHXRBs also show the appearance of strong ($>10$~per~cent fractional rms) and coherent (quality factor $Q\sim$5--10) low-frequency (0.1-10~Hz) quasi periodic oscillations (LFQPOs), at frequencies that correlate with the break frequency during the hard-soft state transitions \citep{KleinWoltvdKlis2008,IngramMotta2019}. These strong LFQPOs, classified previously as Types B and C (e.g. \citealt{Casellaetal2005}) are notably either absent or much weaker in Cyg~X-1. However, recent work has used the cross-spectrum to identify weak (3--6~per cent), low-coherence ($Q\sim$2--3) QPO-like features in NICER data from the soft-hard state transitions of several transient BHXRBs \citep{Mendezetal2024,Alabartaetal2025,Bellavitaetal2025,Brigitte2025}. Such features may correspond to Type C QPOs in this transition \citep{Alabartaetal2025} and have also been observed recently in NICER hard state data from Cyg~X-1 \citep{Koenigetal2024,Fogantinietal2025}. Cyg X-1 approaching and within the soft state shows the reappearance of broadband variability in the form of a singly-bending power-law power-spectral component, with slope $-1$ at low frequencies \citep{axelsson05}. Such a component is also seen in soft state transient BHXRBs but with much lower fractional rms variability than in Cyg~X-1 \citep{Heil15}.

Recently, attention has refocussed on the spectral-timing properties of BHXRBs such as Fourier time lags \citep{nowak99}, which can be used to investigate the relation between disc and coronal variability. Above $\sim2$~keV, Fourier-frequency dependent hard lags (variations at higher photon energies lagging those at lower energies) of $\sim1$--$100$~ms were first discovered in Cyg~X-1 \citep{Miyamotoetal1988} and have been associated with Compton-scattering delays in an extended corona \citep{Kazanasetal1997}, jet or slower outflow \citep{Reigetal2003,KylafisReig2024}, or propagation lags in a radially-extended hot inner accretion flow \citep{kotov01,arevalo06}. However, studies with {\it XMM-Newton EPIC-pn} data showed that, on time-scales of seconds or longer, coronal variations of hard state BHXRBs including Cyg~X-1 lag those from the softest X-rays ($<1$~keV) by up to a few tenths of a second \citep{Uttley11}. These delays are consistent with mass-accretion propagation from the inner disc to a more compact central corona \citep{Uttley11,UttleyandMalzac2025}. Furthermore, on time-scales $<1$~s the sign of the lags changes in several transient BHXRBs, so that the soft band lags behind the coronal variations by up to a few ms \citep{Uttley11,DeMarco2015lags}. These soft lags might be explained in terms of light-travel delays associated with X-ray reverberation of the coronal emission from the disc \citep{Uttley14,Kara19}, or the delay between the rise in coronal seed photons and reverberation, which is also produced by propagation from disc to corona \citep{UttleyandMalzac2025}. 

Models have also been considered where the complex patterns of lags are produced by the multiple additive and overlapping Lorentzian components which also fit the power spectrum, each with their own fixed time-lag or phase-lag \citep{Mendezetal2024,Fogantinietal2025,Bellavitaetal2025}, or more complex lag functions \citep{Jinetal2025}. While such models may provide useful empirical descriptions of frequency-dependent power- and cross-spectra and are helpful to reveal hidden QPO components, their direct physical interpretation is uncertain, because it is known that BHXRB variability on different time-scales is multiplicative, producing the ubiquitous rms-flux relation \citep{Uttleyetal2001,Gleissneretal2004,Heiletal2012} and approximately log-normal flux distributions \citep{Uttley05}. If there are distinct variability processes which can be simply modelled with Lorentzians and these are multiplied together, their Fourier counterparts must be convolved, and then it is difficult to see how such a process can lead to additive and independent power-spectral and cross-spectral features. Multiplicative variability processes are predicted by models of propagating mass-accretion fluctuations \citep{lyu1997,Uttley05}, although such models must be developed further to explain the complex combination of power-spectral and lag behaviour \citep{Rapisardaetal2017,UttleyandMalzac2025}.

Observations of much brighter transient BHXRBs by the {\it Neutron Star Interior Composition Explorer} ({\it NICER}) have shown that the pattern of low-frequency hard and high-frequency soft lags discovered with {\it XMM-Newton} is ubiquitous in hard state BHXRBs \citep{wang22}. The amplitude of high-frequency soft lags also evolves with the QPO frequency and the power-spectral hue \citep{wang22}, a quantity which measures the power-spectral shape \citep[See][and section \ref{sec:power_spec}]{Heil15}. This high-frequency lag evolution could indicate a systematic change in coronal geometry, as the source evolves from the hard to the soft state \citep{Kara19,wang21,UttleyandMalzac2025}. Although Cyg~X-1 does not show evidence for the strong LFQPOs seen in transient outburst rise and hard-soft transitions, the evolution of broadband timing features through the hard state remains very similar to that of transient BHXRBs \citep{Heil15}. Therefore, it is interesting to compare the detailed spectral-timing properties of Cyg~X-1 with transient BHXRBs in the hard state, and ask whether they show similar evolution with changing power-spectral shape.

Up to now, high-frequency soft lags have not been reported for Cyg~X-1, which may be due to limited signal-to-noise compared to the brightest BHXRBs. One problem for recovering the intrinsic spectral-timing properties is that Cyg~X-1 shows significant extrinsic variability on time-scales down to seconds, linked to the clumpy wind of the massive companion star \citep{Grinberg2015,Laietal2024}. However, with a sufficiently long observation which covers orbital phases closer to inferior conjunction (i.e. $\phi_{orb}=0.5$), it is possible to use X-ray colour information to identify time ranges which are free of significant variable absorption, and can be used to study the intrinsic spectral-timing properties in detail, using long exposure times.

In this paper, we perform a comprehensive spectral-timing analysis of data from two long observations of Cyg~X-1 in the hard state. These are the week-long {\it XMM-Newton} observations obtained as part of our 2016 May/June multiwavelength campaign, {\it Cyg~X-1 Hard state Observations of a Complete Binary Orbit in X-rays} (CHOCBOX, see also \citealt{jones21,Lai2022,Laietal2024}), and a 5-day {\it NICER} observing campaign in 2022 May. For comparison, we use {\it NICER} data from the 2018 outburst of one of the best-studied and brightest hard state BHXRBs of recent times, MAXI~J1820+070 (ASASSN-18ey, \citealt{tucker18}), which has been analysed extensively, by e.g. \citet{demarco21,wang21,wang22,Bollemeijer24}. We select MAXI~J1820+070 data from the bright hard state based on the power-spectral hue, to have similar power-spectral shapes to the Cyg~X-1 data, covering hues from $\sim 45^{\circ}$ to $\sim 90^{\circ}$, which is within the hard state \citep{Heil15}. We also use contemporaneous data from the {\it International Gamma-Ray Astrophysics Laboratory} ({\it INTEGRAL}), to allow comparison of spectral-timing properties for the first time over the 0.5--200~keV range, both within a source as hue changes and between sources for similar hue. 

This paper is organised as follows. Section \ref{sec:alldata} introduces the {\it XMM-Newton}, {\it INTEGRAL}, and the {\it NICER} datasets and Section~\ref{sec:results} looks at the spectral-timing data products: energy dependent power spectra, fractional-rms energy spectra and frequency and energy-dependent lags of these data sets. In Section~\ref{sec:discussion} we attempt to provide a consistent explanation of the observed spectral-timing differences between Cyg~X-1 and MAXI~J1820+070 in terms of propagating fluctuation models for the variability.

\section{Observations and data reduction}
\label{sec:alldata}
Our goal in this work is to compare the spectral-timing properties of Cyg~X-1 and MAXI~J1820+070 in the hard state when their power-spectral shapes, as defined by hue (described in Section~\ref{sec:power_spec}), are similar. To cover a range of hue with high-quality data we use the long CHOCBOX {\it XMM-Newton} {\it EPIC-pn} dataset together with a long {\it NICER} observation from May 2022.  We select two {\it NICER} observations of MAXI~J1820+070 with similar hues to the Cyg~X-1 datasets for comparison. We describe the details of the observations and data reduction below.

\subsection{ {\it XMM-Newton}}
\label{sec:xmm-data}
Due to the bright nature of the target and the focus on high-count rate data suitable for spectral-timing analysis, the CHOCBOX {\it XMM-Newton} observations of Cyg X-1 in 2016 were obtained with {\it EPIC-pn} in timing mode as the primary instrument, for a total exposure of $\simeq$ 572 ks over 4 satellite orbits (May 27 - June 2 2016), and we use only the {\it EPIC-pn} data in this work.  We reduce the data using {\it XMM-Newton} Science Analysis software (SAS) v20.0 following the criteria listed in \cite{Lai2022} i.e. selecting the RAWX columns between 30-46, while limiting the PI channels between 500 and 10000, roughly equivalent to 0.5 - 10 keV in the energy space, without any explicit background subtraction, with \texttt{PATTERN == 0}, to consider only single pixel events. \cite{Lai2022} discarded counts from RAWX columns 36-39 to remove pile-up effects on the spectrum, on account of the high source count rates. Although this approach is suitable for spectral analysis, we have verified that including these columns does not significantly affect the spectral-timing behavior while significantly improving the statistical errors on measured time-lags. Therefore we do not exclude these central columns in the detector from our analysis, except for calculation of the time-averaged spectrum which is used in Section~\ref{sec:results}.  The expected background \citep{ng10}, when compared to the source, is both weak and does not vary significantly on the timescales probed by our Fourier analysis.

We use SAS task \texttt{barycen} to obtain the event lists in barycentred time units. The barycentered event list is then converted to a light curve binned at $\simeq3.036$~ms and the light curves are then sliced into $\simeq10$~s segments (3292 bins). The choice of light curve binning is predicated on the timing analysis in the subsequent sections and is set by the {\it EPIC-pn} detector frame time, while the choice of light curve segment size is based on the Good Time Intervals (GTIs) listed in the event list. The telemetry dropouts in the instrument on account of the high source count rates result in GTI segment lengths of $\simeq10$~s spanning over the entire data set. Hence, a segment length is selected such that it again aligns with a multiple of the frame time closest to the average GTI length (9.994512 sec). We retain only segments that contain contiguous data with no gaps (green points in Fig.~\ref{fig:light_curve_Cyg_X1}).

Sections of the dataset affected by stellar wind absorption and manifesting as dips in the light curve are filtered out using spectral colour-colour diagrams \citep{Nowak11, hirsch19, grinberg20} following the methodology listed in \cite{Lai2022}. Light curve segments ($\simeq10$~s) are kept or removed depending on their position in the colour-colour diagram, set by the mean count rates in 3 energy bands. The criteria used are listed in \cite{Lai2022}, i.e segments are filtered out if they have `soft colour' (ratio of 0.5--1.5 keV/ 1.5--3 keV count rates) $\geq$ 0.7 and `hard colour' (ratio of 1.5--3 keV/ 3--10 keV count rates) $\geq0.95$. The resulting diagram showing the selected data is shown in Fig.~\ref{fig:xmm_nicer_cc} (left panel). Once the wind-absorbed sections are removed from the data, about 230 ks of total data are retained, encompassing all four ObsIDs. Following \citep{Lai2022}, we refer to this dataset as {\it XMM-Newton} `No Wind Absorption' (NWA) in table \ref{tab:hue}.

\subsection{{\it NICER}}
\label{sec:nicer_data}
The {\it NICER} data for Cyg~X-1 and MAXI~J1820+070 were obtained from the HEASARC website and reprocessed with the standard task \texttt{nicerl2} with HEASOFT 6.29c \citep{FTOOLS_2014}, released on 31 August 2021, using default parameters, and barycentered using the task \texttt{barycorr}.

{\it NICER} observed Cyg X-1 with the X-ray Timing Instrument (XTI) in the context of an {\it Imaging X-ray Polarimetry Explorer} ({\it IXPE}) campaign in May 2022. In this current work, we present data from ObsId 5100320102--5100320105, obtained on 2022 16-20 May 2022. 
Akin to the {\it XMM-Newton} data, the 0.5--10 keV {\it NICER-XTI} light curve reveals several dips due to stellar wind absorption. We thus construct a colour-colour diagram similar to that obtained for {\it XMM-Newton} data with $\simeq$ 10s light curve segments, shown in Fig.~\ref{fig:xmm_nicer_cc}. Based on the effective removal of the light curve dips, we impose a hard colour $\geq 2.05$ cutoff for retaining the data. The difference in the cut-offs between {\it NICER-XTI} and {\it XMM-Newton EPIC-pn} primarily stem from the differences in the effective area curves of the two detectors, with {\it NICER-XTI} exhibiting a significantly softer response than {\it EPIC-pn}. We have a total of $\simeq$ 69 ks of useful data out of which about $\simeq$ 56 ks is NWA. We refer to this dataset as {\it NICER} 2022 NWA in Table \ref{tab:hue}.

Unlike Cyg~X-1, the MAXI~J1820+070 data sets show no evidence for dipping, commensurate with the low mass of the companion star and only moderately high inclination of the binary orbit \footnote{Estimated between $\simeq$ between $66^{\circ}$ and $81^{\circ}$ \citep{torres20}.} as mentioned in section \ref{section:Intro}. We use $\simeq$ 9 ks of MAXI J1820+070 data taken between April 4-6, 2018 and $\simeq$ 12 ks of data taken on April 21, 2018 at two different states of its spectral-timing evolution spanning over ObsIds 1200120120, 1200120123 and 1200120133, 1200120134 respectively as listed in table~\ref{tab:hue}. We refer to this dataset(s) as {\it NICER} 2018 in table \ref{tab:hue}.   

\subsection{INTEGRAL}
\label{sec:integral_data}
We process {\it INTEGRAL} observations for both Cyg X-1 and MAXI J1820+070 using the Offline Scientific Analysis (OSA) tool within the Docker container\footnote{details on OSA through docker can be found in the URL: \url{https://www.isdc.unige.ch/integral/download/osa/doc/11.2/osa_inst_guide.pdf}} using the steps listed in the IBIS Analysis user manual\footnote{The IBIS science analysis manual can be found in the URL: \url{https://www.isdc.unige.ch/integral/download/osa/doc/11.2/osa_um_ibis.pdf}}. The first step comprises running the task \texttt{ibis\_science\_analysis} to select IBIS/ISGRI events from levels COR to DEAD, within the entire Science Window (ScW). Subsequently, we run the task \texttt{ii\_pif}, to determine the Pixel Illumination Fractions for events from a particular source (Cyg X-1 or MAXI J1820+070) within the field of view, followed by the task \texttt{evts\_extract} with the flag \texttt{barycenter} set to 1, to obtain events from that source in barycentered time units. We then select events from the list with the condition \texttt{PIF\_1==1} as recommended by the IBIS science analysis manual and also in \cite{potts06conf}, to maximise signal-to-noise. The INTEGRAL-ISGRI (20--200 keV) data sets sometimes show periodic dropouts every 4 or 8 seconds which can be attributed to the inherent cycle duration of data packets in the ISGRI system\footnote{\url{http://isdc.unige.ch/integral/support/faq.cgi?IBIS-014}}. The dropouts occur when the telemetry capacity dedicated to ISGRI reaches its limit, resulting in the periodic omission of certain data packets from transmission to the ground station. This behavior is visible in MAXI J1820+070 but is not discernible in Cyg~X-1 data, likely due to the count-rate differences between the two sources. For spectral-timing analysis, we consider the interruptions in ISGRI telemetry as invalid time intervals which are removed by excluding them from the GTIs. We utilise all GTIs $\geq$ 8 seconds for timing analysis ensuring a low-frequency cutoff at $\simeq$ 0.125 Hz for MAXI J1820+070.   

Cyg X-1 observations with {\it INTEGRAL} as a part of the CHOCBOX campaign were taken between May 31, 2016, and June 2, 2016, with ScW-IDs 1684 and 1685 as listed in Table \ref{tab:hue} totaling $\simeq$ 334 ks of coverage. The observations of the same source between 2022 May 16, 2022, and May 20, 2022, with concurrent coverage from {\it NICER}, have $\simeq$168~ks of data with ScW-IDs 2304 and 2305. MAXI J1820+070 {\it INTEGRAL} observations between April 4-6 2018 correspond to ScW-IDs 1938 and 1939 with $\simeq$36~ks of data, while the observation on April 21, 2018 corresponds to ScW-ID 1944 with $\simeq$41~ks of data.

\begin{figure*}
    \centering
    \includegraphics[width=\textwidth]{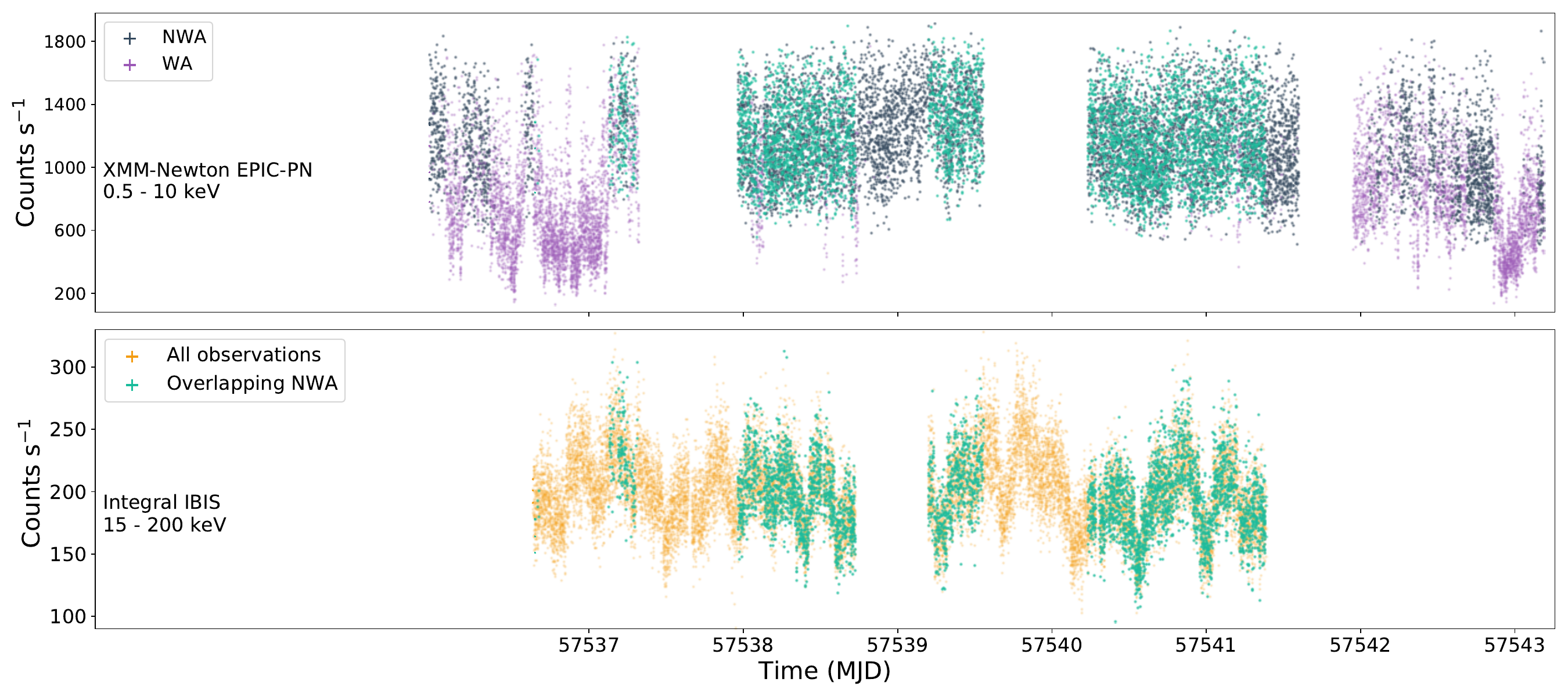}
    \caption{Cyg~X-1 {\it XMM-Newton EPIC-pn } (0.5--10 keV) and {\it INTEGRAL} (15--200 keV) light curves, thinned to show only every 20th datapoint for easy visualisation. NWA (No wind absorption) and WA (Wind absorption) intervals in the {\it XMM-Newton} light curve have been annotated by asphalt and purple respectively. In the lower panel, the full extent of {\it INTEGRAL} observations have been annotated in yellow, while the segments in green across both panels denote NWA sections of {\it XMM-Newton} data with simultaneous coverage from {\it INTEGRAL}. {\it XMM-Newton} covering starts at 57535.9 MJD, while {\it INTEGRAL} coverage starts at MJD 57536.7. }
    \label{fig:light_curve_Cyg_X1}
\end{figure*}

\subsection{Simultaneous 0.5--10~keV and {\it INTEGRAL} data}
In this work (Section~\ref{sec:lags}) we present the first {\it joint} spectral-timing analysis of 0.5--10~keV ({\it XMM-Newton} or {\it NICER}) and {\it INTEGRAL} data, to investigate the energy-dependent lags over a broad energy range. To do so, we must use only simultaneous (in barycentred time) data from both observatories, obtained from overlapping GTIs for the instruments used while, for the case of Cyg~X-1, only using times when the data is identified as NWA according to {\it XMM-Newton} or {\it NICER} colours. For example, the {\it XMM-Newton} (0.5 - 10 keV) and {\it INTEGRAL} (15 - 200 keV) light curves (with $\simeq 10$~s sampling) are shown in Fig.~\ref{fig:light_curve_Cyg_X1}, highlighting the wind-absorbed and NWA sections of the data. The green data points in both panels denote the NWA segments which are simultaneous to both light curves and are used for broadband spectral-timing analysis in Section~\ref{sec:lags}. We repeat this approach to obtain simultaneous {\it NICER} and {\it INTEGRAL} data. The data points in Fig.~\ref{fig:light_curve_Cyg_X1} have been thinned by a factor of 20, for easy visualisation. The long term count rate variations ($\simeq$ hours timescales) in the {\it INTEGRAL} light curve visible in the lower panel of Fig.~\ref{fig:light_curve_Cyg_X1}, are artifacts of observations taken within each SCW and do not reflect the variability of the source at these timescales. 

\begin{figure*}
    \includegraphics[width=0.9\columnwidth]{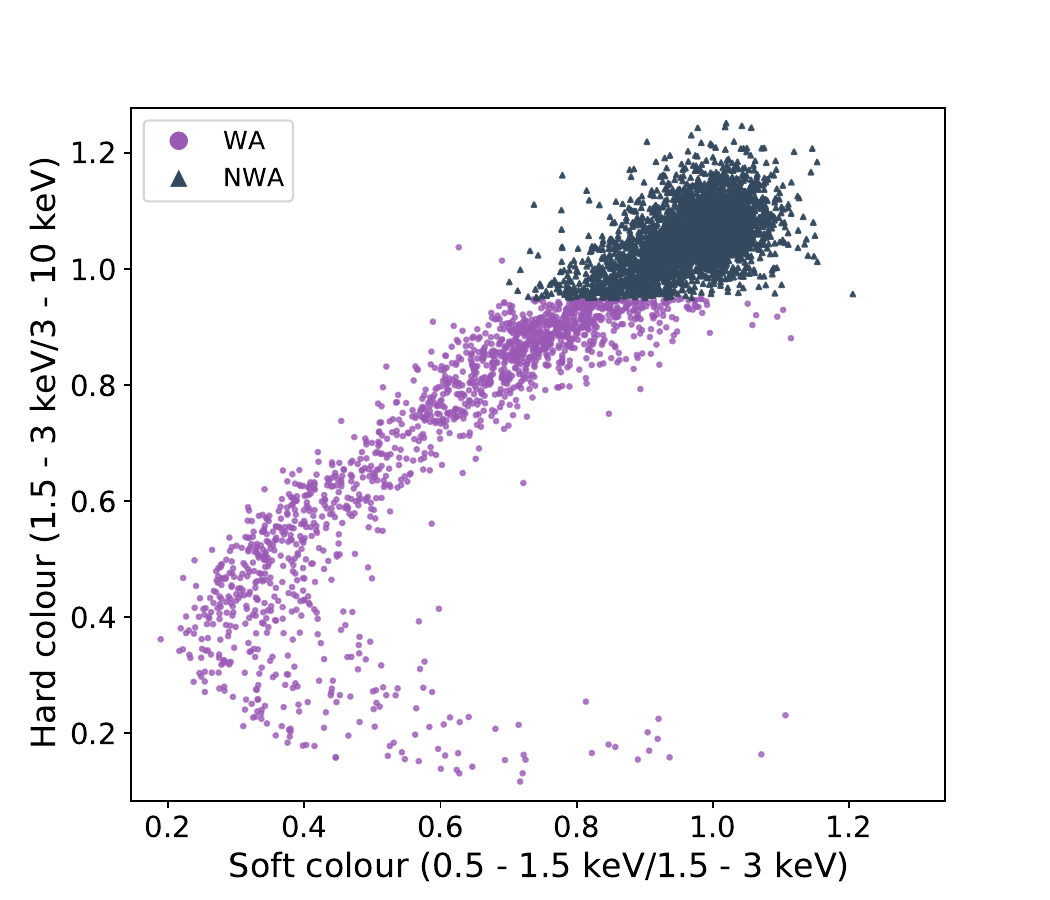} 
    \includegraphics[width=0.9\columnwidth]{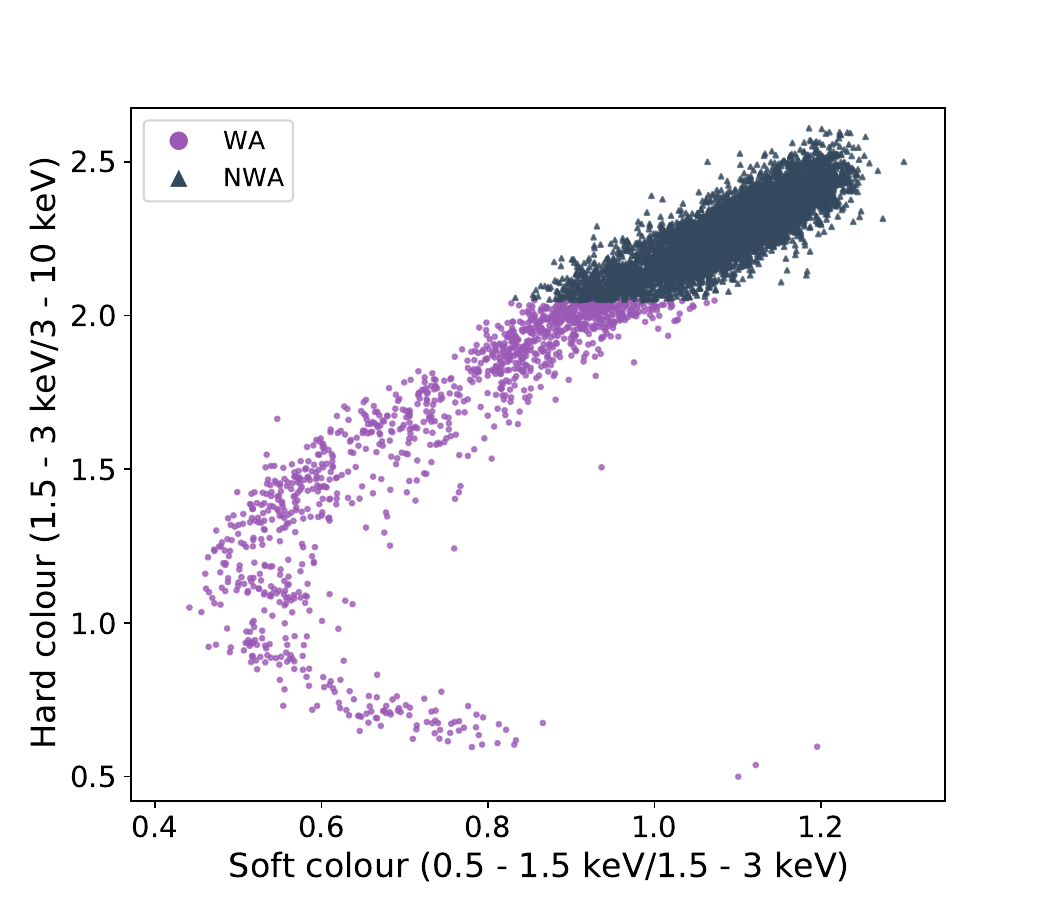}
    \caption{Cyg~X-1 colour-colour diagrams for 10~s segments of the {\it XMM-Newton EPIC-pn} (left) and {\it NICER} (right) datasets before filtering on wind absorption. Similar to Fig.~\ref{fig:light_curve_Cyg_X1}, NWA (No wind absorption) and WA (Wind absorption) intervals have been annotated using the same colours. Wind-absorbed light curve segments defined as showing hard colour $<$ 0.95 and 2.05 for {\it EPIC-pn} and {\it NICER} respectively, and soft colour $<0.7$ (for both instruments) are removed from the light curves used in this work.}    
    \label{fig:xmm_nicer_cc}
\end{figure*}

\section{Results}
\label{sec:results}

\begin{figure}
    \centering
    \includegraphics[width=0.47\textwidth]{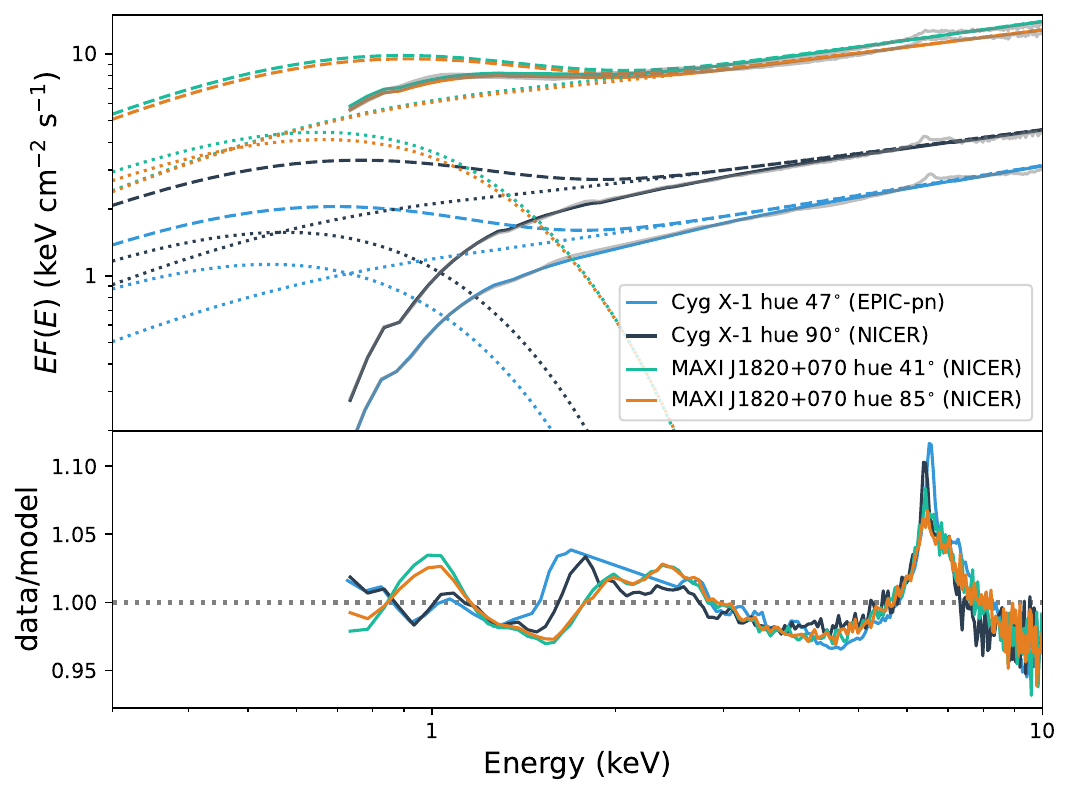}
    \caption{Comparison of X-ray spectra and data/model ratio for the four datasets considered here (data plotted in grey). The solid lines show the unfolded spectral models obtained from fitting simple Comptonized disc blackbody models to the data in the 0.7--10~keV range, accounting for neutral absorption. The dashed and dotted lines show respectively the unabsorbed model and model components (disc blackbody and Comptonized spectrum). Error bars have been removed for clarity and the data rebinned to bins of width 50~eV for plotting purposes.}
    \label{fig:spectrum_compare}
\end{figure}

\subsection{Semi-empirical spectral fits}
Our focus in this paper is on spectral-timing analysis and comparison between Cyg~X-1 and MAXI~J1820+070 at different hues in the hard state. However, it is informative to first compare their X-ray flux density spectra, which we show in Fig.~\ref{fig:spectrum_compare}. Following the approach of \citet{Koenigetal2024}, we use a simplified prescription to model the spectra by including a 5~per~cent systematic error in our model fit to allow for spectral complexity associated with X-ray reflection while obtaining a semi-empirical parameterisation of the spectral shape. We fit the data in the 0.7--10~keV range, excluding 1.7--2.5~keV in the {\it XMM-Newton} data, to remove the regions containing the uncertain Si and Au edge features and the Charge Transfer Inefficiency (CTI) feature in the {\it EPIC-pn}  instrument response. We fit our semi-empirical model using {\sc xspec v12.14} and consists of a disc blackbody ({\sc xspec} model {\sc diskbb}, \citealt{Mitsudaetal1984}) with some fraction of disc photons upscattered into a thermal Comptonization component ({\sc thcomp}, \citealt{Zdziarskietal2020}), all absorbed by neutral Galactic absorption ({\sc tbabs}, \citealt{Wilmsetal2000}). The neutral absorbing columns of Cyg~X-1 and MAXI~J1820+070 have been well-studied with H {\sc i} surveys and/or more detailed spectral-fitting and so we fix their values respectively at $7\times10^{21}$~cm$^{-2}$ \citep{HI4PIcollab2016} and $1.4\times10^{21}$~cm$^{-2}$ \citep{Kajavaetal2019}.   

Although the specific fit parameters are not directly meaningful and we do not record them in detail here, they do enable a comparison of the relative shapes of the X-ray spectra. We find that the photon index $\Gamma$ is similar for the four datasets, with values of $\Gamma=1.59$ and $\Gamma=1.67$ for the respectively low and high-hue {\it XMM-Newton} and {\it NICER} observations of Cyg~X-1, and $\Gamma=1.65$ and $\Gamma=1.68$ for the equivalent {\it NICER} observations of MAXI~J1820+070. The true indices must be steeper once reflection is accounted for, but their similarity indicates a common spectral state for all four datasets. The inferred disc temperatures for Cyg~X-1 ($kT\simeq0.23$~keV and 0.24~keV for low and high hue respectively) and MAXI~J1820+070 ($kT\simeq 0.28$~keV in both datasets) and their strengths relative to the Comptonized continuum are also fairly similar, consistent with these systems showing similar accretion disc geometry. 

It is notable that there is a factor 3--5 difference in flux between Cyg~X-1 and MAXI~J1820+070. Given the radio parallax distances of Cyg~X-1 ($2.22^{+0.18}_{-0.17}$~kpc, \citealt{jones21}) and MAXI~J1820+070 ($2.96\pm 0.33$~kpc, \citealt{Atrietal2020}), the integrated $0.01-1000$~keV unabsorbed model luminosities are $3(2.5)\times10^{37}$~erg~s$^{-1}$ and $3.7(3.3)\times10^{37}$~erg~s$^{-1}$ for Cyg~X-1 (low and high hue) and $2.1(1.8)\times10^{38}$~erg~s$^{-1}$ and $1.8(1.6)\times10^{38}$~erg~s$^{-1}$ for MAXI~J1820+070 (low and high hue). Here we assume a thermal Comptonization cut-off energy of 100~keV,  with values assuming a cut-off of 50~keV shown in parentheses. The range of cut-off energies is chosen based on prior measurements in Cyg~X-1 \citep{Basaketal2017} and MAXI~J1820+070 \citep{Buissonetal2019} respectively. Thus, for the observations studied, the luminosity of MAXI~J1820+070 is likely up to 6 times greater than that of Cyg~X-1. We will discuss these results in the context of our spectral-timing comparison in Section~\ref{sec:discussion}.

Since the spectra are relatively similar and thus span a narrow range of hardness on a hardness-intensity diagram (HID), we do not show their location on the HID here. However, for comparison with the equivalent colour-luminosity diagram shown by \cite{Koenigetal2024} (their Fig.~1), Cyg~X-1 shows unabsorbed model 1.5--10~keV luminosity relative to Eddington luminosity, $L_{\rm 1.5-10~keV}/L_{\rm Edd}=1.5\times10^{-3}$ and $2.3\times10^{-3}$ for low and high hue respectively, with corresponding colours\footnote{Defined by \citet{Koenigetal2024} in terms of 1.5--5~keV and 5--10~keV unabsorbed fluxes as $(F_{\rm 5-10~keV}-F_{\rm 1.5-5~keV})/(F_{\rm 5-10~keV}+F_{\rm 1.5-5~keV})$.} -0.09 and -0.13 (the black hole masses assumed for the fractional luminosities are discussed in Section~\ref{sec:discussion_bhmass}). MAXI~J1820+070 shows $L_{\rm 1.5-10~keV}/L_{\rm Edd}=0.037$ and $0.035$ for low and high hue respectively, and corresponding colours -0.13 amd -0.14. These values place the observations in the hard state locus in terms of colour and on the expected track for Cyg~X-1 but with MAXI~J1820+070 hard state fractional luminosity significantly elevated compared to Cyg~X-1.

\begin{table*} 
\centering
\begin{tabular}{lccccc} 
\hline  
\textbf{Dataset (Source)} &  \textbf{PC1} &  \textbf{PC2} & \textbf{Hue} & \textbf{State} & \textbf{$\langle \tau_{\rm M-S} \rangle$ (ms)}\\
\hline 
\textbf{Cyg X-1} \\
\hline 

\textbf{Data details}&INTEGRAL ScW-Ids & \multicolumn{3}{|c|}{1684, 1685 and 2304, 2305} \\
& {\it XMM-Newton EPIC-pn} ObsIds & \multicolumn{4}{|c|}{0745250201, 0745250501, 0745250601, 0745250701} \\

& {\it NICER-XTI} ObsIds & \multicolumn{4}{|c|}{5100320102, 5100320103, 5100320104, 5100320105} \\
\hline

{\it INTEGRAL (XMM-Newton) } 2016 NWA & 4.59$\pm$0.1  &  0.76$\pm$0.01 &  46.8$\pm$2.3$^{\circ}$  &  Low hue hard state & 1.3$\pm$0.6\\

\hline
{\it NICER (INTEGRAL)} 2022 NWA & 10.41$\pm$0.34 & 1.06$\pm$0.02 & 89.6$\pm$1.3$^{\circ}$ &  High hue hard state & N/A\\
&  (11.28$\pm$0.61) & (1.03$\pm$0.1) & (93.17$\pm$5.6$^{\circ}$)  & & \\

\hline
\textbf{MAXI J1820+070}\\
\hline 

\textbf{Data details}&INTEGRAL ScW-Ids & \multicolumn{3}{|c|}{1938, 1939 and 1944} \\

& {\it NICER-XTI} ObsIds & \multicolumn{4}{|c|}{1200120120, 1200120123 and 1200120133, 1200120134} \\
\hline

{\it NICER (INTEGRAL)} 2018 & 4.11$\pm$0.32 & 1.74$\pm$0.05 & 40.9$\pm$3.3$^{\circ}$ &  Low hue hard state & 1.71$\pm$0.08\\
\hline
{\it NICER (INTEGRAL)} 2018 & 11.55$\pm$0.64 & 1.36$\pm$0.03 & 85.4$\pm$2.0$^{\circ}$ &  High hue hard state & 1.00$\pm$0.09\\
\hline

\end{tabular}

\caption{Data description for {\it XMM-Newton}, {\it NICER} and {\it INTEGRAL}, along with PC, Hue and $\langle \tau_{\rm M-S} \rangle$ values for MAXI J1820+070 and Cyg X-1 at both high and low hue. The second row in the 2022 NWA data refers to the PC and hue values for {\it INTEGRAL} 20 -- 200 keV data to show consistency between the {\it NICER} 4.8 -- 9.6 keV and the  {\it INTEGRAL} 20 -- 200 keV bands.} 
\label{tab:hue}
\end{table*}

\subsection{Broadband power spectra and power colours}
\label{sec:power_spec}
The spectral-timing analysis of the reduced datasets was performed using \texttt{Python 3.7} \citep{python3} scripts developed following the methodology listed in \cite{Uttley14}, while simultaneously utilising certain wrapper classes from \texttt{Stingray} v2.1 \citep{stringray-1}. 

To compare the hard state spectral-timing properties of Cyg~X-1 with those of the well-studied transient source MAXI~J1820+070, we first classify the power-spectral shape using the power colours and power-spectral `hue' approach of \citet{Heil15}. This method requires us to calculate the ratios of integrated power in four geometrically-spaced frequency bands from 1/256~Hz to 16~Hz. The power colour and `hue' classification, while empirical, offers a model-independent method for comparing timing evolution with other observables \citep{Heil15,wang22}. The low-frequency band requires contiguous light curve segments of at least 256~s, which are not available for the 2016 {\it XMM-Newton} data due to telemetry dropouts (see Section~\ref{sec:xmm-data}). Therefore we use {\it INTEGRAL} 20--200 keV data to determine the power colours for the 2016 data. The segment size is $\simeq256$~s for power-colour measurement from the power spectrum, while the time bin size remains the same as for the {\it XMM-Newton} data i.e. $\simeq3$~ms, for consistency. For the remaining Cyg X-1 and MAXI~J1820+070 data we use the NICER hard band (4.8--9.6~keV) light curve binned at 1/256s, which yields power-colour measurements\footnote{We have checked the consistency of the hue values obtained from the {\it NICER} 2022 and {\it INTEGRAL} 2022 NWA Cyg~X-1 datasets and the {\it INTEGRAL} 20--200 keV band serves as a proxy for the {\it NICER} 4.8 -- 9.6 keV bands (see Table \ref{tab:hue}).} consistent with those measured from 2--13~keV {\it Rossi X-ray Timing Explorer} ({\it RXTE}) Proportional Counter Array (PCA) data by \citet{Heil15}. 

The high-frequency Poisson noise is estimated by fitting a constant to the power spectrum at frequencies $>100$~Hz which is then subtracted from the entire power spectrum \citep{Uttley14}. The 100~Hz lower bound is necessitated by the lower temporal-resolution of {\it EPIC-pn} data, and we also use it for the {\it NICER} data for consistency. Note that with extensive and high-quality data, intrinsic source variability in the hard state can be detected at 100 Hz, and to frequencies up to $\sim300$~Hz with a power-law slope of $<-2$ (e.g. for Cyg~X-1 see \citealt{Revnivtsevetal2000}, see also Fig.~5 of \citealt{Fogantinietal2025}). This intrinsic noise will contaminate the Poisson noise-level estimate from the fitted ranges by a small amount, which we estimate to be $<5\times10^{-6}$~Hz$^{-1}$ in fractional rms-squared power density at energies above 2~keV (and substantially smaller at lower energies, where the intrinsic variability power is suppressed at high frequencies). The contamination produces a small negative bias in measured fractional rms-squared which is between 0.5 and 2~per~cent of the inferred intrinsic variance in the highest (2--16~Hz and 8--32~Hz) frequency bands used for power colour and fractional rms spectra calculations (this Section and Section~3.4). The bias is much smaller at lower frequencies. Such a bias is negligible compared to the statistical errors and source and frequency/energy dependent differences in spectral-timing properties shown in this work.

\begin{figure*}
    \centering
    \includegraphics[width=0.85\textwidth]{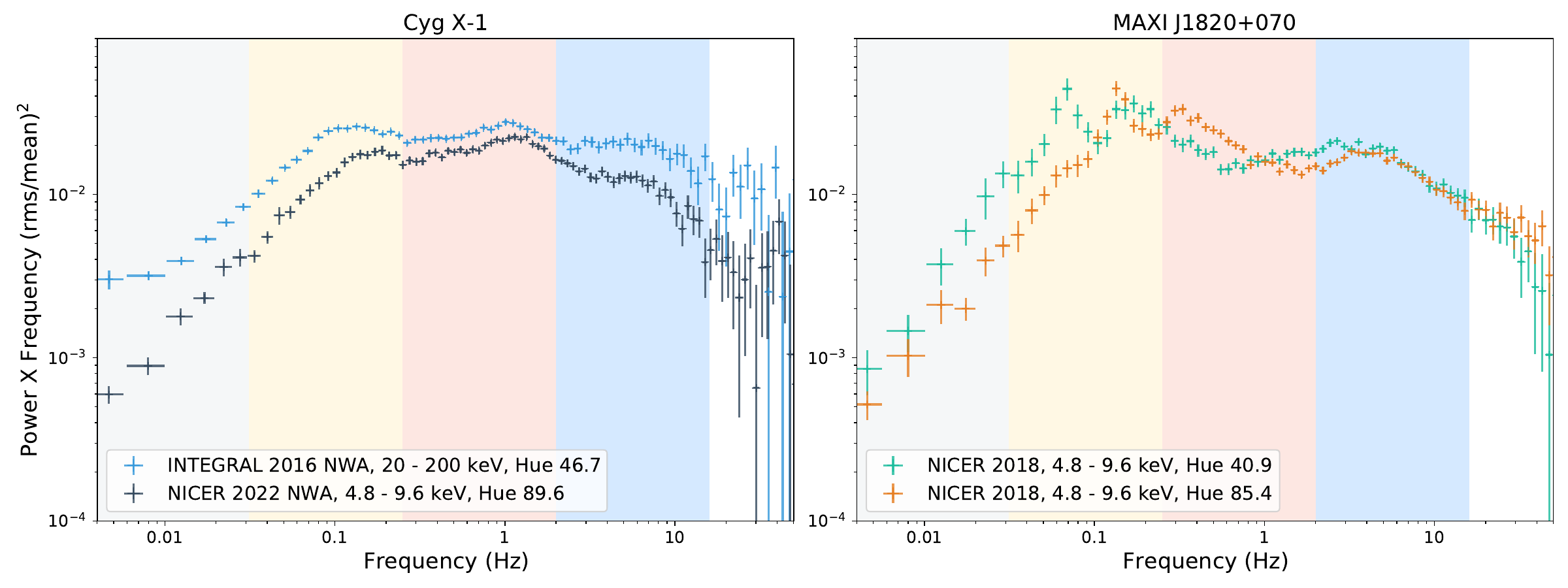}
    \caption{Power spectra plotted in power $\times$ frequency units used for power-colour calculations for the four datasets listed in table \ref{tab:hue}. Left (Cyg X-1): {\it INTEGRAL} 2016 NWA (20 - 200 keV) at hue 46.7$^{\circ}$ and {\it NICER} 2022 NWA (4.8 - 9.6 keV) at hue 89.6$^{\circ}$. Since the {\it INTEGRAL} power-spectral amplitude is suppressed by the background contribution, it has been scaled up by a factor of 10 in the figure to aid visual comparison of the Cyg~X-1 power spectra. Right (MAXI J1820+070): {\it NICER} 2018 (4.8 - 9.6 keV) at hues 40.9$^{\circ}$ and 85.4$^{\circ}$. The four broad frequency bands: 1/256 -- 1/32~Hz, 1/32 -- 0.25~Hz, 0.25--2~Hz, and 2--16 Hz used to calculate power colours and then hue are annotated in grey, yellow red, and blue respectively. }
    \label{fig:pc_ps_all}
\end{figure*}

Fig.~\ref{fig:pc_ps_all} shows the power spectra used to determine the power colours of the source for each observation; along with the four broad frequency bands: 1/256 -- 1/32~Hz, 1/32 -- 0.25~Hz, 0.25--2~Hz and 2--16 Hz used to calculate power colours annotated in grey, yellow red and blue respectively. Note that the {\it INTEGRAL} 20--200 keV power spectral amplitude is artificially suppressed by photon flux associated with background and other sources in the field-of-view,  which cannot be removed from the event lists but are not expected to vary significantly on the time-scales probed (see also \citealt{potts06conf}). Since the power colours are ratios of integrated power in different frequency bands, they remain unaffected by this dilution effect. However, to aid visual comparison with the {\it NICER} Cyg~X-1 data, the {\it INTEGRAL} power spectrum in Fig.~\ref{fig:pc_ps_all} is scaled up by a factor of 10.

\begin{figure}
    \centering
    \includegraphics[width=0.47\textwidth]{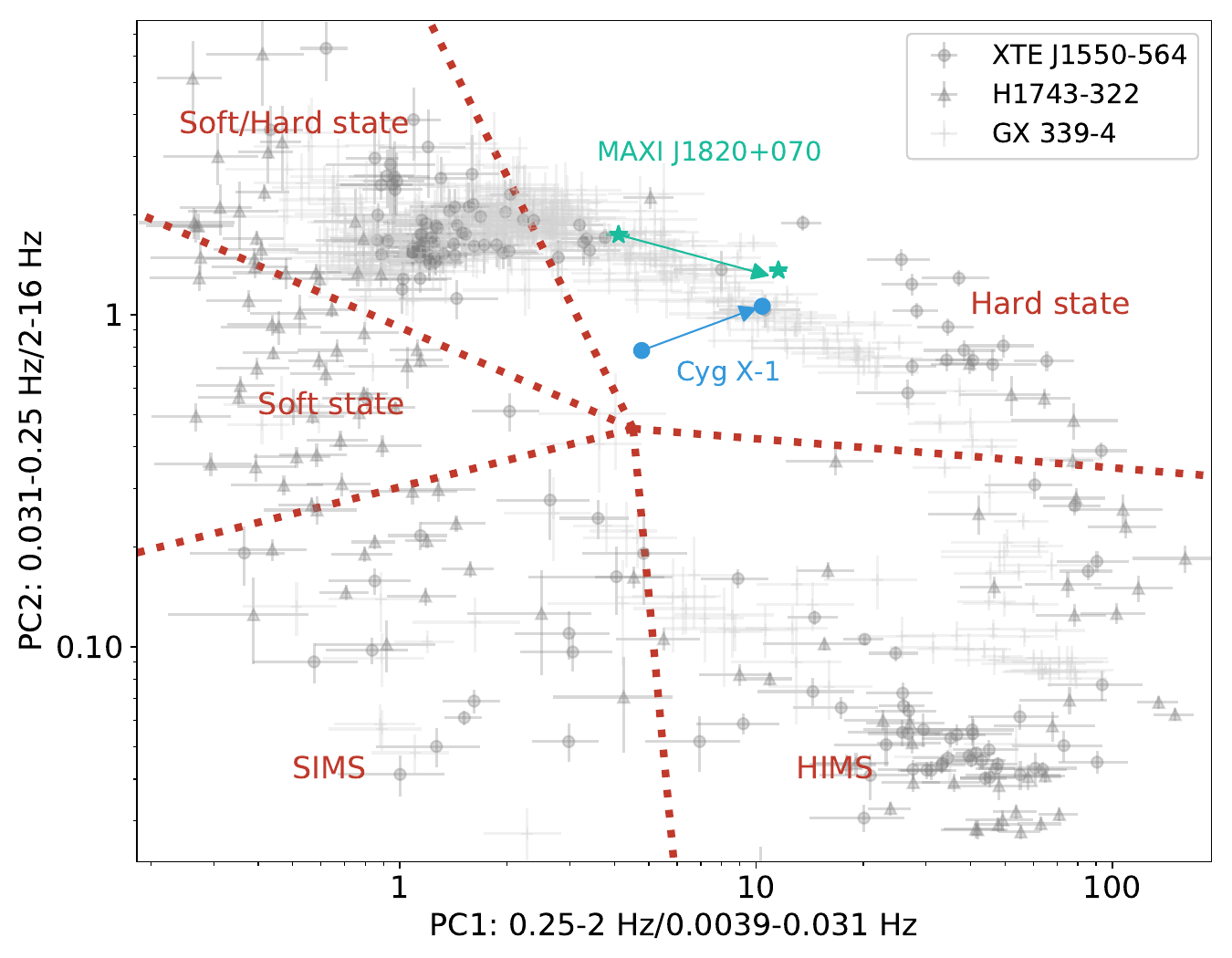}
    \caption{$PC1$ and $PC2$ values for Cyg X-1 and MAXI J1820+070. The green/blue arrows show the evolution of both sources from hues $\simeq 41-45^{\circ}$ to $\simeq 85 - 89^{\circ}$. For comparison the grey data points in the background show the evolution of three transient BHXRB sources - XTE J1550-564, H1743-322 and GX 339-4 - across states (data from \citealt{gardenier18}).}
    \label{fig:pc_compare}
\end{figure}

We use the noise-subtracted power spectra to generate power-colour ratio 1 ($PC1$), the ratio of integrated powers in 0.25--2~Hz/(1/256)--(1/32) Hz (ratio of red/grey bands in Fig.~\ref{fig:pc_ps_all}) and power-colour ratio 2 ($PC2$), the ratio of integrated powers in (1/32)--0.25~Hz/2--16 Hz (ratio of yellow/blue bands in Fig.~\ref{fig:pc_ps_all}) as mentioned in \cite{Heil15}. Based on a large sample of BHXRBs which followed a characteristic elliptical evolution on the $PC1$-$PC2$ plane, \cite{Heil15} defined the centre of the power-colour-colour diagram $(PC1_{0},PC2_{0})=(4.5192, 0.4537)$. Defining that point as the origin, any point on the $PC1$-$PC2$ plane can be described by a position vector $[\log(PC1/PC1_{0}),\log(PC2/PC2_{0}]$, the argument of which (measured clockwise from the position vector $[-1,1]$) is called the hue. With these definitions, power spectra characteristic of the hard state correspond to hue values from $340^{\circ}$--$140^{\circ}$. The hue values for the power spectra shown in Fig.~\ref{fig:pc_ps_all} are shown in the corresponding figures and are listed in Table \ref{tab:hue}. Note that while different hue values are seen for each dataset for a given source and/or instrument, all power spectra correspond to the hard state and moreover, the MAXI~J1820+070 data are chosen so that their power spectra show similar hue values to those of the two Cyg~X-1 datasets. 

Fig \ref{fig:pc_compare} shows the $PC1$ and $PC2$ values for both sources across hues plotted on the power-colour-colour (PCC) diagram \citep{Heil15}. The green/blue arrows show the evolution of MAXI J1820+070/Cyg X-1 from hues $\simeq 41-45^{\circ}$ to $\simeq 85 - 89^{\circ}$. The spectral states in the PCC diagram, defined following \citep{Bellonietal2005} as hard, hard intermediate (HIMS), soft intermediate (SIMS) and soft, are classified according to power-colours following \cite{Heil15}. For comparison, the grey data points show the power-colour evolution (based on {\it RXTE} PCA data) of three transient BHXRB sources - XTE J1550-564, H1743-322, and GX 339-4 - taken from \cite{gardenier18}. 

Note that Cyg~X-1 power-colours reported by \citep{Heil15} show a broader track than seen for individual transient BHXRBs, which could explain the somewhat anomalous location of the lower hue data-point for Cyg~X-1 compared to MAXI~J1820+070. This difference in $PC2$ may also relate to clear differences between the detailed shapes of the power spectra of Cyg~X-1 and MAXI~J1820+070 in Fig.~\ref{fig:pc_ps_all}, including three vs. two broad 'humps' in Cyg~X-1 compared to MAXI~J1820+070 and the presence (at frequencies around 0.1~Hz) of a low-frequency QPO and its harmonic in MAXI~J1820+070. However, Fig.~\ref{fig:pc_ps_all} also highlights that the similar hues for both sources are driven primarily by the similarity of their low-frequency power-spectral breaks, which predominantly determine PC1 and the value of hue along the hard state track.

\subsection{Energy-dependent power spectra}
For calculating the energy-dependent power spectra and fractional rms spectra we use the {\it XMM-Newton EPIC-pn} data in $\simeq10$~s segments filtered for wind absorption (NWA), sampling a frequency range $\simeq0.1$--$164$~Hz. The choice of segment size for the {\it NICER} data is constrained by the $\simeq$10 second segment length in the {\it XMM-Newton} 2016 Cyg X-1 data, while also limiting the loss of light curve segments within GTIs that contain effects of wind absorption. We use 16 sec light curve segments binned at 1/256 sec such that it encompasses the variations in 0.0625--128 Hz. We again estimate the high frequency noise by fitting a constant to the power spectra at frequencies $\geq$ 100 Hz.

\begin{table}
\centering
\begin{tabular}{llc}
\hline
\textbf{Instrument}          & \textbf{Energy Range (keV)} & \textbf{Band Name} \\
\hline
{\it XMM-Newton / NICER}           & 0.5 -- 1                   & S                  \\
                              & 2.5 -- 4.0                   & M                \\
                              & 4.8 -- 9.6                   & H                \\
\hline                              
{\it INTEGRAL}                     & 20 -- 80                   & H1               \\
                              & 80 -- 200                  & H2               \\
\hline                              
\end{tabular}
\caption{Energy ranges and corresponding band names for {\it XMM-Newton}, {\it NICER}, and {\it INTEGRAL}.}
\label{tab:instrument_bands}
\end{table}

\begin{figure*}
    \centering
    \includegraphics[width=0.85\textwidth]{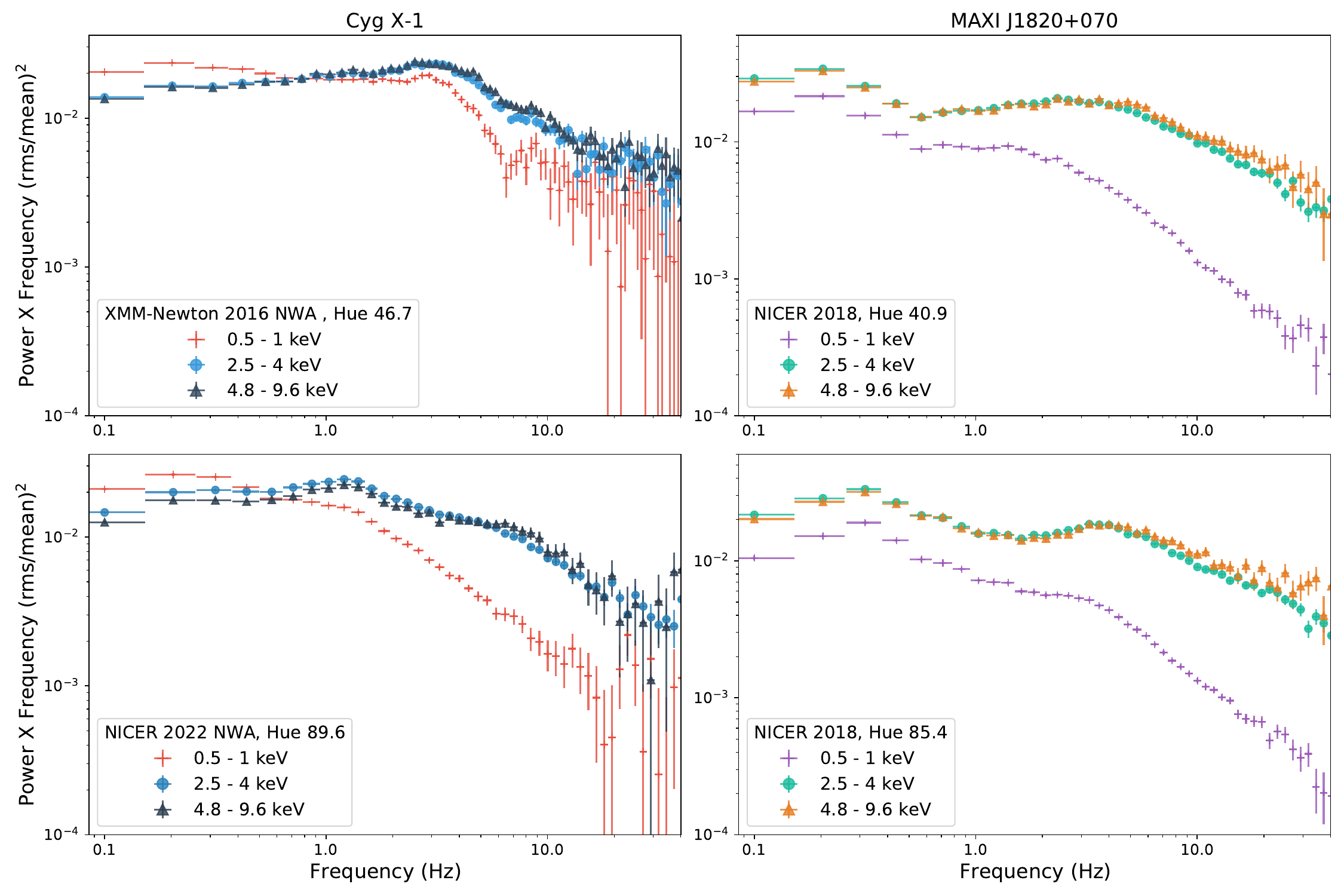}
    \caption{Energy-dependent power spectra for the four datasets listed in Table \ref{tab:hue} for 0.5--1 keV (S), 2.5 -- 4.0 keV (M) and 4.8 -- 9.6 keV (H) energy bands. M and H band power spectra are similar, but the high-frequency variability amplitudes increase as one moves to higher energy bands, while the S band variabilities are suppressed at these frequencies for both sources.}
    \label{fig:xmm_ps_comb}
\end{figure*}

Fig.~\ref{fig:xmm_ps_comb} shows the power spectra for the {\it XMM-Newton EPIC-pn} and {\it NICER} data we consider here, for energy bands 0.5--1 keV (soft band S, also where disc emission becomes significant), 2.5 -- 4.0 keV (medium-band, M), 4.8 -- 9.6 keV (hard band, H), as listed in table \ref{tab:instrument_bands} We do not consider the {\it INTEGRAL} data since dilution by the high background level makes the normalisation difficult to interpret. The well-known increase in high-frequency variability amplitude as one moves to higher-energy bands \citep{nowak99,grimberg14} is seen for Cyg X-1. The effect is more subtle between M and H bands for MAXI~J1820+070, but can be discerned once the small normalisation differences in the power spectra are accounted for. The S band variability is significantly suppressed towards higher frequencies in all cases, as also noted by \citet{Wilkinsonetal2009} for hard state {\it XMM-Newton} data for BHXRBs GX~339-4 and SWIFT~J1753.5-0127 (see also \citealt{Cassatellaetal2012}), later for Cyg~X-1 by \cite{Lai2022} and also for {\it NICER} data for MAXI J1820+070 in the S and M bands \citep{wang21}. However, at low frequencies, Cyg~X-1 shows enhancement of the S band variability compared to the higher-energy bands, while for MAXI~J1820+070 the S band variability amplitude remains lower than that of the harder bands across the entire $\simeq$ 0.1-32~Hz frequency range. We also note that the Cyg~X-1 2022 higher-hue hard state shows S band variability that falls below the higher-energy variability amplitudes at significantly lower frequencies than seen in the 2016 hard state data. Any differences between the MAXI~J1820+070 S band power spectra appear to be less substantial than those seen in Cyg~X-1.

\subsection{Fourier-resolved fractional rms spectra}

\begin{figure*}
    \centering
    \includegraphics[width=0.85\textwidth]{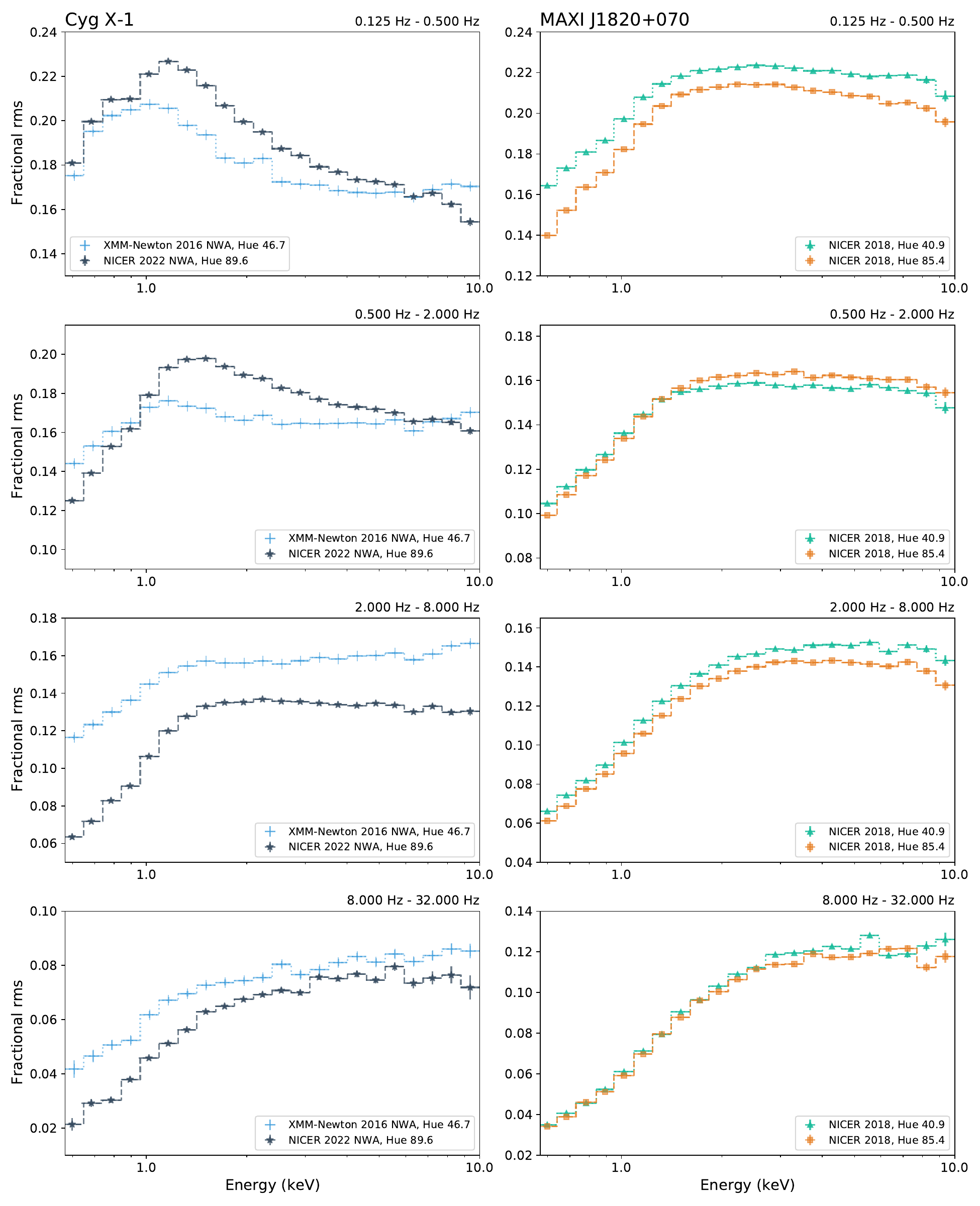}
    \caption{0.5--10~keV rms spectra for frequency ranges 0.125--0.5~Hz, 0.5--2~Hz, 2 --8 Hz, and 8--32 Hz calculated for both the Cyg X-1 (left) and MAXI J1820+070 (right) across different hues. The rms-energy spectra of Cyg~X-1 appear to be more different in shape across hues when compared to the differences seen for MAXI~J1820+070, across the 0.125 -- 32~Hz frequency range.}
    \label{fig:rms_comb}
\end{figure*}

The energy-dependence of the power spectra can be interpreted with the help of the Fourier-resolved fractional rms spectrum (e.g. see \citealt{Revnivtsevetal1999,Uttley14}), which plots as a function of energy the fractional rms calculated for a given frequency range for smaller energy bins ($\simeq$ 24 in the 0.5 -- 10 keV range) than used for the power-spectral plots. To make fractional rms spectra, we use the same segment size and time binning as the energy-dependent power spectra. Fig.~\ref{fig:rms_comb} shows the 0.5--10 keV Poisson-noise subtracted fractional rms spectra for four frequency ranges: 0.125--0.5~Hz, 0.5--2~Hz, 2--8~Hz, and 8--32~Hz for both Cyg X-1 (left) and MAXI J1820+070 (right) and qualitatively illustrates the simultaneous relative variations at the S band and harder energies in different frequency ranges. As was the case for the energy-dependent power spectra, we do not include the {\it INTEGRAL} data due to the dilution effects of the background which will likely also be energy-dependent. The rms spectra suggest that the distinctive behaviors of the soft band power spectra for both Cyg~X-1 and MAXI~J1820+070 may be linked to the disc, since a clear break in rms behavior (either a fall or a rise) can be seen at and below the highest energies ($\sim2$~keV) where hard state disc emission is expected (e.g. see also \citealt{Wilkinsonetal2009,Cassatellaetal2012}).

For Cyg~X-1 at lower frequencies, there is a pronounced increase in variability within the S band, reaching $\simeq$ 20\% in rms amplitude across both hues. Conversely, as we ascend to higher frequencies, the rms amplitude decreases significantly to $\simeq$ 4-6\%, emphasizing the heightened variability of the S band at lower frequencies, while the S band rms diminishes relative to the harder emission at higher frequencies. In the case of MAXI J1820+070, the S band rms is lower than that of the rms at harder energies across all frequency ranges and both hues, re-emphasizing the results in Fig.~\ref{fig:xmm_ps_comb} (top and bottom-right). Overall, the rms spectra of Cyg~X-1 appear to be more distinct in shape for the different hues, compared to the differences seen for MAXI~J1820+070, at both high and low frequencies, which follow a virtually identical trend for both hues.

\subsection{Lag-frequency spectra}
\label{sec:lags}
\begin{figure*}
    \centering
    \includegraphics[width=0.8\textwidth]{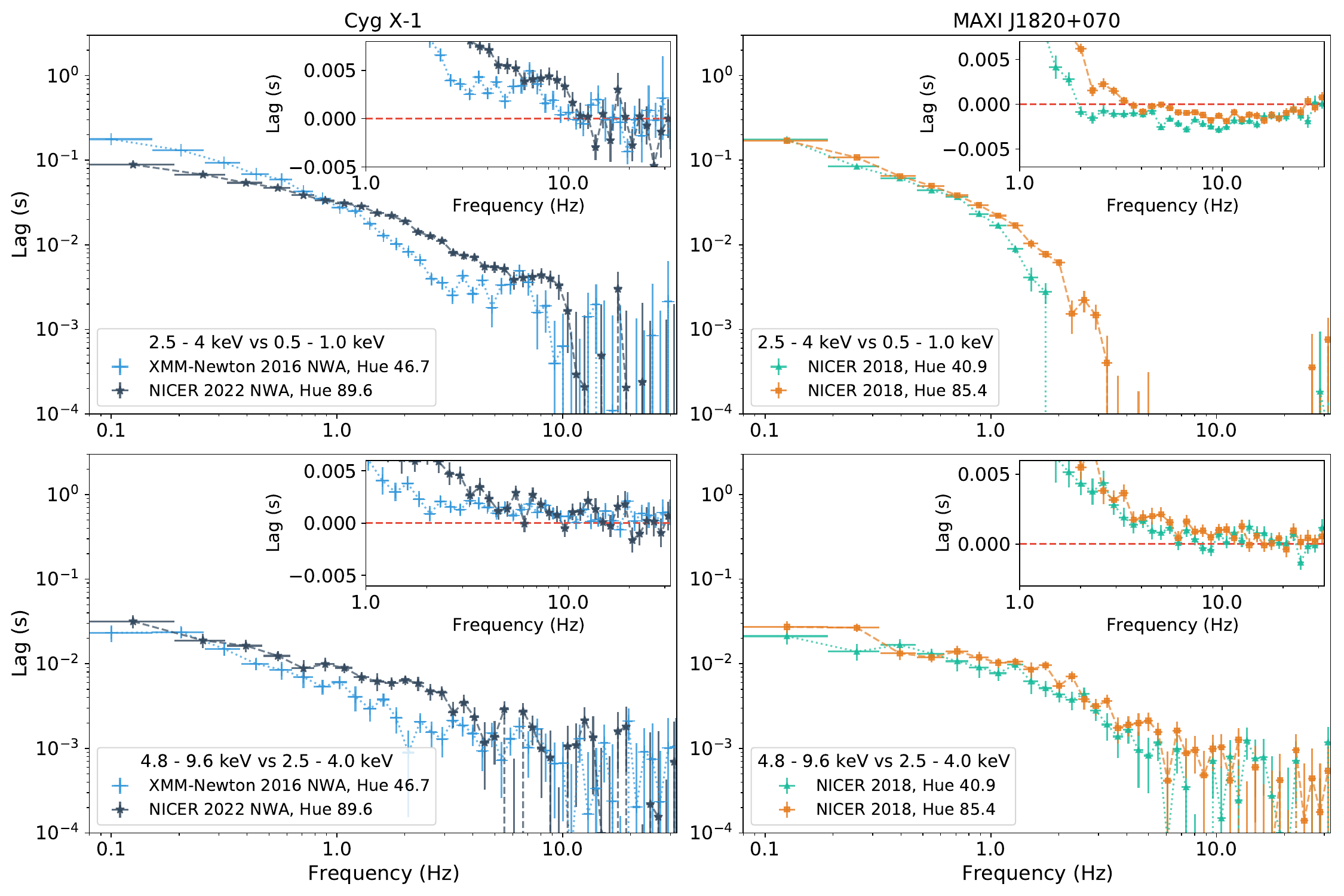}
    \caption{Top left(right): 2.5--4~keV vs. 0.5--1~keV (M-S) lag frequency spectra for both the Cyg X-1 (MAXI J1820+070) datasets at different hues. Bottom left (right): 4.8--9.6~keV vs. 2.5--4~keV (H-M) lag frequency spectra for both Cyg X-1 (MAXI J1820+070) datasets at different hues. The insets show the zero crossing frequency, where the lags cross from negative to positive (if at all) as well as any negative, `soft' lags for frequencies $\simeq$1-- 32~Hz in all four panels on a linear scale, while the outer panels highlight the low-frequency ($\leq$ 1~Hz) hard lags for both sources across hues on a logarithmic scale.}
    \label{fig:xmm-nicer-lf}
\end{figure*}

We use the four datasets to generate lag-frequency spectra (following the prescription in \citealt{Uttley14} and see \citealt{Miyamotoetal1988,nowak99} for earlier applications to Cyg~X-1 data), using the same segment size, time, and energy binning as for the energy-dependent power spectra. The top panels of Fig.~\ref{fig:xmm-nicer-lf} compare the 2.5--4~keV vs. 0.5--1~keV (M-S) lag-frequency spectra for each source at different hues, while the lower panels compare the 4.8--9.6~keV vs. 2.5--4~keV (H-M) lag-frequency spectra. The insets highlight the zero crossing frequency, where the lags cross from negative to positive (if at all) as well as any negative, `soft' lags for frequencies $\simeq$1-- 32~Hz in all four panels. 

The ubiquitous low-frequency `hard' lags are observed in both sources for M-S and H-M cases along with the characteristic decrease in lag amplitude as a function of frequency. As seen in many sources \citep{Uttley11,grimberg14,DeMarco2015lags,demarco21}, the M-S lags are significantly steeper than the lags between harder energy bands which show the approximate $\tau \propto \nu^{-0.7}$ diminution for both hues, albeit with the possibility of distinct steps at certain frequencies for Cyg~X-1 \citep{nowak99,Nowak2000,misra17}. At higher frequencies ($\geq$ 2 Hz) M-S `soft' lags in the order of 1-1.5 ms are discernible for MAXI J1820+070 (see Fig \ref{fig:xmm-nicer-lf} inset on the upper right). The M-S soft lags in the lag-frequency spectrum are not significant for Cyg~X-1, as evidenced in Fig.~\ref{fig:xmm-nicer-lf} (top-left), despite the high quality of data provided by the long {\it XMM-Newton} and {\it NICER} observations. 

The mean soft lag amplitude $\langle \tau_{\rm M-S} \rangle$, calculated using the methodology of \cite{wang22} for M-S lags in Cyg X-1 is $1.3\pm0.6$~ms at the lower hue, while for MAXI J1820+070 the mean soft lags are 1.71 $\pm$ 0.08 ms at the lower hue and $1.00 \pm 0.09$~ms at higher hue. $\langle \tau_{\rm M-S} \rangle$ could not be estimated for Cyg~X-1 at the higher hue, due to the lack of even moderately significant negative lag values.  Superficially, the soft lags derived using the method of \citet{wang22} are similar between sources. However these lags are estimated based on the frequency ranges where the lags go negative in the lag-frequency diagram, and those ranges are very different: 19--36~Hz in Cyg~X-1 and 2--22~Hz and 3.2--22~Hz for the low and high hue MAXI~J1820+070 observations. A visual comparison of the insets of both upper panels in Fig.~\ref{fig:xmm-nicer-lf} reveals a clear difference, with highly-significant negative M-S lags appearing in the 1--10~Hz frequency range in MAXI~J1820+070 while the corresponding lags for Cyg~X-1 are clearly positive.

The M-S lags for Cyg~X-1 exhibit different frequency-dependent behavior for both hues, characterized by a steeper drop in lag amplitudes at higher frequencies from a higher initial amplitude at low frequencies at the lower hue, in contrast to a more gradual diminution with frequency starting from a lower initial amplitude at the higher hue. The M-S lags for MAXI J1820+070 evolve identically across both hues before diverging close to the frequency where the transition to negative 'soft' lags occurs.

\begin{figure*}
    \centering
    \includegraphics[width=0.8\textwidth]{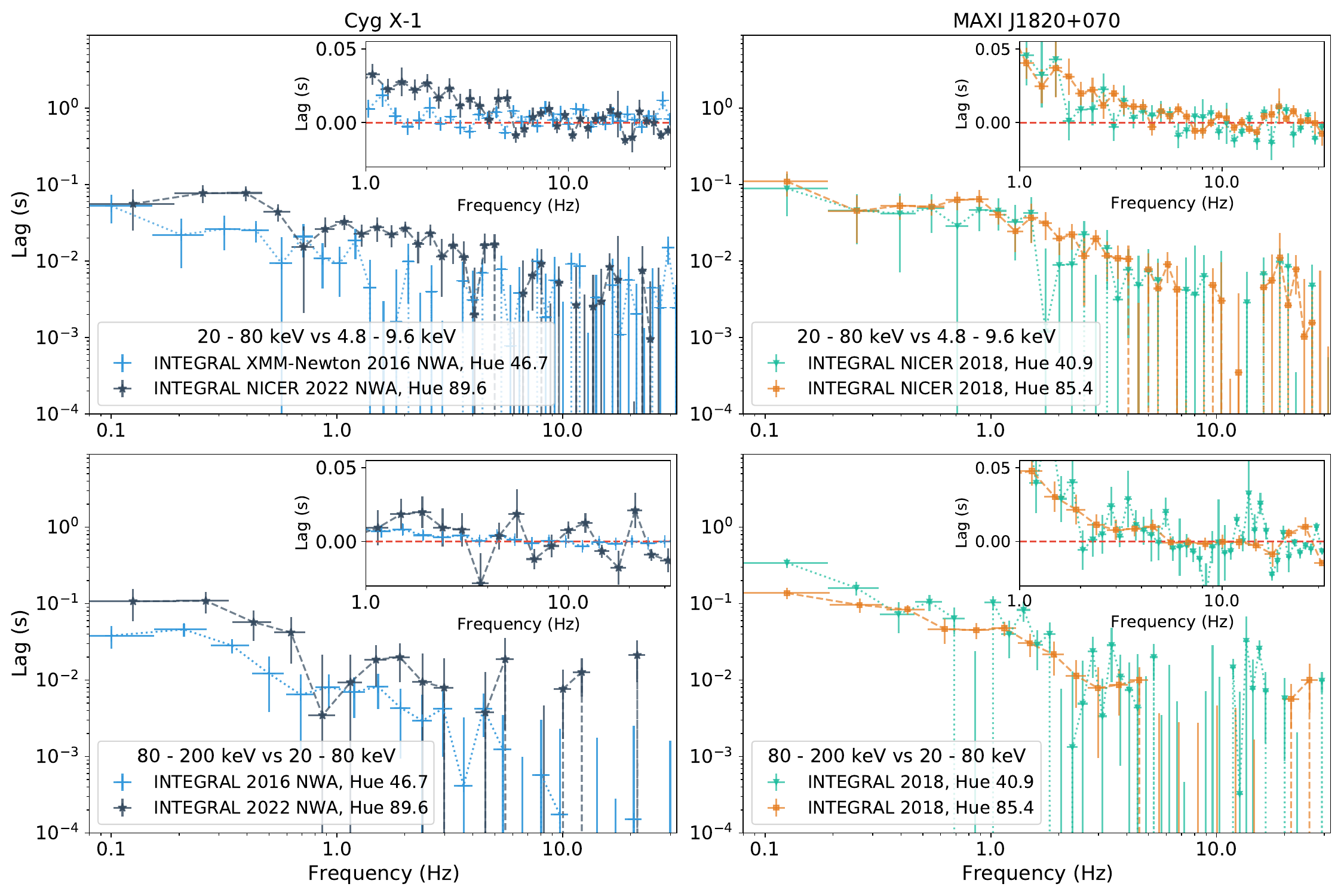}
    \caption{Top left (right): 20--80~keV vs 4.8--9.6~keV (H1-H) lag frequency spectra for Cyg X-1 (MAXI J1820+070) at different hues, Bottom left (right): 80 - 200 keV vs 20 - 80 keV (H2-H1) lag frequency spectra for Cyg X-1 (MAXI J1820+070) at different hues. The insets highlight the zero crossing frequency, if the lags cross from negative to positive, for frequencies $\simeq$1-- 32~Hz in all four panels on a linear scale, while the outer panels highlight the low-frequency ($\leq$ 1~Hz) hard lags for both sources across hues on a logarithmic scale}
    \label{fig:xmm_nicer_lf_pl}
\end{figure*}

Fig.~\ref{fig:xmm_nicer_lf_pl} shows the {\it INTEGRAL} 20--80~keV (H1) vs. 4.8-9.6~keV (H) lag vs. frequency for both sources in the top panels. Lag vs. frequency between the {\it INTEGRAL} 80--200~keV (H2) and 20--80~keV (H1) bands are shown in the lower panels. The {\it INTEGRAL} band names and corresponding energy ranges are listed in table \ref{tab:instrument_bands}. We have limited the segment size to 8 s for the MAXI J1820+070 {\it INTEGRAL} event lists due to telemetry issues as mentioned in section \ref{sec:integral_data}, to strike a balance between the lowest frequency being probed (length of the segment) and the observed uncertainties, that depend on the number of available segments being averaged. Both the H1-H and H2-H1 lags show trends similar to that of H-M lags in Fig.~\ref{fig:xmm-nicer-lf} (bottom left and bottom right) i.e. a drop in lag amplitude as a function of frequency. 


\subsection{Lag-energy spectra}
Following the prescription in \citet{Uttley14}, we generate lag-energy spectra using the same segment and time bin size, frequency ranges, and energy binning as the fractional rms spectra. The common reference band for the measurement of lags for individual energy bins (channels-of-interest, CI) is 0.5--10 keV, to maximize signal-to-noise ratio. We use the same reference band to calculate lags from simultaneous {\it INTEGRAL} data and extend the lag-energy spectra from 0.5--200~keV. The reference band for CIs in the range 0.5--10.0 keV requires subtraction of the CI band light curve from the 0.5--10 keV light curve, to avoid contamination of those lags by correlated observational noise \citep{Uttley14}. 

\begin{figure*}
    \centering
    \includegraphics[width=0.85\textwidth]{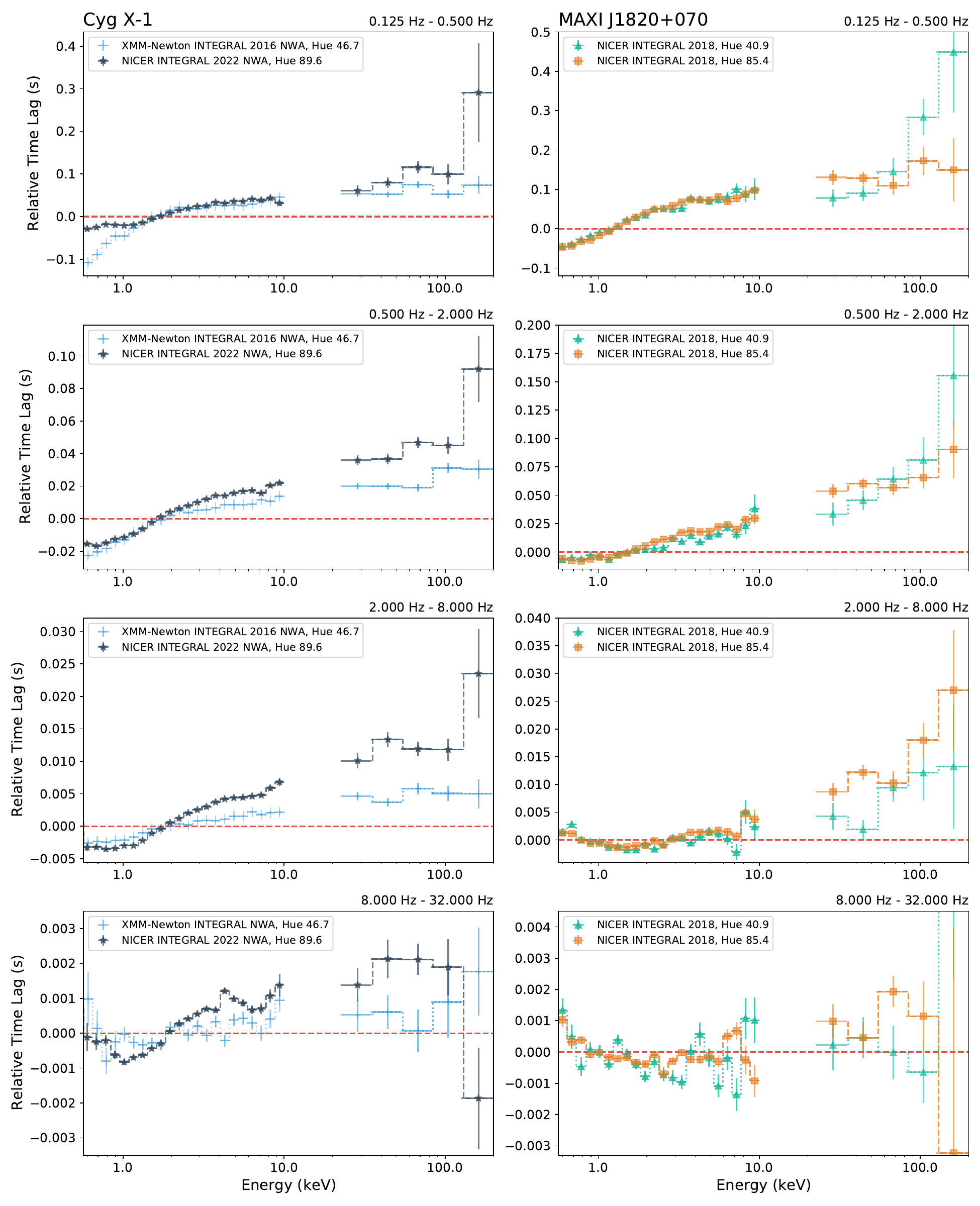}
    \caption{0.5--200 keV lag energy spectra, plotted for frequencies 0.125--0.5~Hz, 0.5--2~Hz, 2--8~Hz and 8--32~Hz as observed in Cyg X-1 (left) and MAXI J1820+070 (right) across different hues. At lower frequencies, above $\sim2$~keV the lags follow an approximately log-linear energy dependence to $\sim100$~keV, but the relation steepens $<2$~keV (with additional complexity at lower energies in the Cyg~X-1 higher-hue data). At higher frequencies, the steep lag-energy gradient in soft X-rays flattens and then inverts, as seen in the M-S lag vs. frequency spectra in Fig.~\ref{fig:xmm-nicer-lf}. Depending on the zero-crossing frequency of the lags, the soft lags only become visible in the 8--32 Hz band for Cyg~X-1 at the lower hue and the 2--8 Hz and 8--32 Hz bands for MAXI~J1820+070 across both hues.}
    \label{fig:tl_comb}
\end{figure*}

Fig.~\ref{fig:tl_comb} shows the combined 0.5--200~keV lag-energy spectra for the four sets of frequency ranges for Cyg~X-1 (left) and MAXI~J1820+070 (right). At energies $>2$~keV the lag vs. energy dependence follows the well-known, roughly log-linear, dependence of lag with energy \citep{nowak99,kotov01,Uttley11}. It is interesting to note, however, that there appears to be a steepening of the high-energy (70 -- 200 keV) lag vs. energy dependence for MAXI~J1820+070 at the lower hue in the lowest frequency range. We tested the significance of this steepening by comparing the quality of the fit of a single log-linear law versus a broken log-linear law, which accommodates the possible steepening at high energies. We fitted data only $>3$~keV (since additional complexity is seen at lower energies) and found a reduction in $\chi^{2}$ statistic of 15.7 for two additional model parameters: the high energy slope and the break, found to be $60\pm13$~keV. This result suggests that the steepening is significant at $> 3\sigma$ significance. However, we do not detect such a significant steepening in the lags in the other datasets, so we note it as interesting for further study but do not speculate further on this feature here. 

The Cyg~X-1 high-energy lag amplitudes are higher at higher hue across all frequencies when compared to the ones at lower hue, as also seen in the H-M lag frequency spectra. At energies $<2$~keV however, we see more complex behaviour in the low-frequency ($<2$~Hz) lag-energy spectra, notably an inflection to a steeper gradient with flattening at still lower energies visible in some cases (e.g. Cyg~X-1 NICER data). These soft X-ray lag behaviours were originally noted by \citet{Uttley11} who attributed them to the effects of propagation delays through the accretion disc, which begins to contribute significantly to the emission below $<2$~keV.

At higher frequencies, the steep lag-energy gradient at low energies begins to flatten and then inverts, corresponding to soft lags as seen in the M-S lag vs. frequency spectra in Fig.~\ref{fig:xmm-nicer-lf}. Depending on the zero-crossing frequency of the lags, this effect becomes visible in the 8--32 Hz band for Cyg~X-1 at hue 46.7$^{\circ}$ and in the 2--8 Hz and 8--32 Hz bands for MAXI~J1820+070 across both hues. For Cyg~X-1, this represents the first known detection of soft lags at high frequencies, although the weakness and the low energies of this upturn prevent soft lags from being seen in the broader-band M-S lag vs. frequency spectrum of Cyg X-1. Note that for Cyg~X-1 we cannot currently detect a possible reflection signature lag \citep{kotov01, Kara19} around the Fe~K emission region with high confidence\footnote{We checked this using finer energy bins between 5-8 keV while simultaneously probing higher frequencies up to $\simeq$80 Hz following \cite{Kara19}.}. This lack of detectable reflection delay (and the small/absent soft lag in Cyg~X-1) is consistent with the relatively small disk-corona separation ($\sim9$~$R_{g}$) inferred from reverberation lag modelling of the reflection signature in Cyg~X-1 by \citet{Mastroserioetal2019}.

\section{Discussion}
\label{sec:discussion}

We have compared high-quality spectral-timing data from Cyg~X-1 at different points in the hard state (defined in terms of the power-spectral evolution, via the hue), and found notably different behaviour when compared to MAXI~J1820+070 at similar hues. The main differences are as follows:
\begin{enumerate}
\item The power spectra and rms spectra (fig. \ref{fig:xmm_ps_comb} and {\ref{fig:rms_comb}}) show that both the amplitude and hue-dependence of the S band variability in Cyg X-1 are different from that of MAXI~J1820+070. Cyg X-1 shows soft X-ray variability which is more prominent than higher-energy variability at low frequencies, and which increases with increased hue (the effect is most notable in the rms-spectra). In contrast, MAXI~J1820+070 shows high-energy variability dominating over the whole frequency range, and only weak dependence on hue.
\item The M-S low-frequency lag behaviour (fig. \ref{fig:xmm-nicer-lf}) also shows much stronger hue-dependence in Cyg~X-1 compared to MAXI~J1820+070, with the lower-hue data showing a steeper lag-frequency dependence than at higher hue in Cyg~X-1, while in MAXI~J1820+070 only the lags above 1~Hz differ between hues. In the lag-energy spectra, (fig. \ref{fig:tl_comb}) these differences manifest in the Cyg~X-1 higher-hue data as a 'filling in' of the usual down-turn which is still seen at energies $<2$--$3$~keV in MAXI~J1820+070 and in the Cyg~X-1 lower-hue data.
\item High-frequency soft lags are not seen in lag-frequency spectra of Cyg~X-1, although upturns in the 8-32~Hz lag-energy spectra can be seen below 1~keV (fig. \ref{fig:tl_comb}). While most significant in the higher-hue hard state, this upturn appears to fill in a down-turn $<3$~keV, so does not lead to significant negative lags in the lag-frequency spectrum. MAXI~J1820+070 shows the well-studied soft lags in the 2--10~Hz frequency range, with higher-hue showing a smaller soft lag and higher crossing-frequency, as shown by \citet{Kara19,wang22}.
\item Lags between bands $\geq2$~keV (fig. \ref{fig:xmm_nicer_lf_pl}) are different between hues in Cyg~X-1. The higher-hue hard state shows steeper lag vs. energy dependence $>2$~keV compared to the lower-hue data (fig.~\ref{fig:tl_comb}), resulting in larger lags. For MAXI~J1820+070 the H-M lag differences are much less pronounced, although clearer differences are seen in the {\it INTEGRAL} bands in fig.~\ref{fig:tl_comb}.
\end{enumerate}
To explain these differences, we consider the physical differences between both systems at the time of observation, specifically the black hole mass, accretion disc temperature and spectral-shape, and the accretion geometry or specific accretion state.

\subsection{Differences due to black hole mass}
\label{sec:discussion_bhmass}
\citet{wang22} has shown that a sample of BHXRBs observed by {\it NICER} show similar evolution of their high-frequency soft lags with power-spectral hue, suggesting a common evolution of accretion geometry. Cyg~X-1 shows no evidence of soft lags except at the lowest energies in the 8--32~Hz lag vs. energy diagram. Its lag vs. frequency behaviour shows lags consistent with zero above 10~Hz, and clearly positive lags in the 4--10~Hz range, in contrast with the negative lags in that range seen for MAXI~J1820+070. However, Cyg~X-1 has a significantly higher estimated black hole mass ($21.2\pm2.2$~$M_{\odot}$, \citealt{jones21}) compared to MAXI~J1820+070 ($6.75^{+0.64}_{-0.46}$~$M_{\odot}$, \citealt{Mikolajewskaetal2022}), suggesting that the time-scales being compared are not equivalent. Lowering the frequency ranges used for power-colour and hue measurements of Cyg~X-1 by a factor 3 (consistent with the higher mass black hole) leads to changes of hue from 47$^{\circ}$ to 148$^{\circ}$ and from 90$^{\circ}$ to 168$^{\circ}$ for the lower- and higher-hue hard state respectively. This would move Cyg~X-1 closer to the HIMS where the BHXRB soft lags reported by \citet{wang22} are larger than those seen in the hard state. Furthermore, the soft time lags (and frequency range for soft lags) in Cyg X-1 should also be scaled by mass, so we should expect even larger soft lags and at lower frequencies than observed in MAXI~J1820+070. Thus, correcting for the mass scaling of variability time-scales only increases the differences in lag behaviour between Cyg~X-1 and MAXI~J1820+070.

\subsection{Differences due to accretion disc temperature and spectrum}
The clearest differences in spectral-timing behaviour between the hard state observations of Cyg~X-1 and MAXI~J1820+070 presented here, are in the M-S lags and also the rms spectra at low energies. The large differences in soft X-ray timing properties suggest a connection to the disc emission. \citet{UttleyandMalzac2025} present a model which explains both hard and soft lags in terms of mass-accretion fluctuations which propagate through the disc before reaching the corona. The low-frequency M-S hard lags are explained by the propagation delay through the blackbody emitting region of the disc to the power-law emitting corona. The high-frequency M-S soft lags can be explained by propagation-induced variations in seed photons from the disc {\it preceding} the coronal heating (and corresponding disc reverberation) caused by the same propagating signal. This effect causes lower-energy coronal power-law variations to precede the reverberation emission from the disc, but with a delay much longer than the light-travel time from the corona to the disc. 

In the Appendix to their paper, \citet{UttleyandMalzac2025} show that the model-predicted hard and soft M-S lags are sensitive to the choice of disc energy band relative to the maximum of the radially-dependent disc temperature, $kT_{\rm max}$ (see their Fig.~C1). Since higher-energy disc photons (relative to the disc maximum temperature) preferentially originate from smaller radii, their propagation-related delays are significantly smaller than for lower-energy photons. Therefore, we might expect to see much smaller or non-existent M-S soft lags in Cyg~X-1 if the energies probed (0.5--1~keV) are larger relative to $kT_{\rm max}$ than in MAXI~J1820+070, i.e. $kT_{\rm max}$ in Cyg~X-1 should be significantly lower than in MAXI~J1820+070. Such a difference might be expected due to the higher observed luminosity and lower black hole mass in MAXI~J1820+070 compared to Cyg~X-1. Using unabsorbed broadband (0.01--1000~keV) fluxes estimated from the simple model fits shown in Fig.~\ref{fig:spectrum_compare}, together with the estimated distance and black hole mass for each system, we estimate that the accretion rate $\dot{m}$, relative to the Eddington limit (and assuming similar radiative efficiency in both sources) is up to 19 times larger in MAXI~J1820+070 than in Cyg~X-1. For a standard accretion disc and identical truncation radius (in $R_{g}$), $kT_{\rm max}\propto M_{\rm BH}^{-1/4}\dot{m}^{1/4}$, which would imply a factor $\sim2.8$ times lower disc temperature in Cyg~X-1 compared to MAXI~J1820+070. Such a difference could in principle explain the absence of soft lags according to the model of \citealt{UttleyandMalzac2025}. 

However, it is important to note that observationally, there is little evidence for such a large difference in disc temperature between the two systems at the observed times and their respective luminosity levels. The crude semi-empirical model fit shown in Fig.~\ref{fig:spectrum_compare} indicates only $\sim17$~per~cent higher disc temperature in MAXI~J1820+070 compared to Cyg~X-1. Even taking the lower value ($kT_{\rm max}\simeq 0.19$~keV) obtained by \citet{Lai2022} for the Cyg~X-1 {\it EPIC-pn} data from a full reflection fit including a soft Comptonizing component, the temperature reduction compared to MAXI~J1820+070 is less than 50~per~cent, compared to the factor $\sim2.8$ lower temperature that is expected based on accretion rate and black hole mass differences. We are left to assume that in fact, the hard state disc emission observed in Cyg~X-1 and/or MAXI~J1820+070 does not behave as expected from a standard multi-temperature blackbody-emitting disc. For example, their spectral hardening factors \citep{ShimuraTakahara1995} may also be different from each other, or the inferred disc temperatures may be distorted due to the presence of the additional steep-spectrum and low-temperature hot plasma component included by \citet{Lai2022}, e.g. if it corresponds to a warm Comptonizing layer above or close to the disc (e.g. see \citealt{Yamadaetal2013} and \citealt{MahmoudDone2018}). Such effects could be integrated into future spectral and spectral-timing model comparisons of Cyg~X-1 and other sources.

\subsection{Differences due to accretion geometry and state}
Since the disc spectral-timing properties likely depend on the disc inner radius, the different behaviours of Cyg~X-1 and MAXI~J1820+070 may be linked to different inner disc radii. For example, in the model of \citep{UttleyandMalzac2025}, the low-frequency break in the power spectrum corresponds to the outermost disc radius where significant accretion variability is generated, which may only correspond to a few tens of gravitational radii. However, the lags and power spectra at high frequencies depend on the inner radius where the accretion flow transitions from disc to corona. If this radius is systematically smaller in Cyg~X-1 compared to MAXI~J1820+070, it might explain the lack of high-frequency soft lags and the relatively high disc temperature in Cyg~X-1 (given its mass and accretion rate). However, a smaller disc radius would imply a different balance of disc and coronal power, which seems contrary to the similar spectral shapes of both sources at these hues. Such a difference in disc radius might alternatively be linked to a significantly lower (or retrograde) black hole spin for MAXI~J1820+070 \citep{Zhaoetal2021_1820,Diasetal2024}, compared with the near-maximal spin inferred for Cyg~X-1 by \citet{Zhaoetal2021_CygX1}\footnote{Note however that deviations from a disc blackbody spectrum due to a warm Comptonizing layer can significantly lower the inferred spin of Cyg~X-1, \citealt{Zdziarskietal2024}}. However, this latter interpretation would imply that the disc extends to the innermost stable circular orbit, meaning that the corona must be produced via some other mechanism than a hot inner accretion flow, which would raise further questions over the formation of the hard coronal power-law emission and production of lags. One would also need to explain the similarity of MAXI~J1820+070 lag evolution to a number of other transient BHXRBs \citep{wang22}, which would seem unlikely if MAXI~J1820+070 has atypically low spin.

Finally, it is interesting to note the possibility that the low and high luminosity versions of the hard state seen respectively in these observations of Cyg~X-1 and MAXI~J1820+070 may be intrinsically different in their spectral-timing behaviour, despite showing similar hue and spectral properties. In a recent paper, \citet{Koenigetal2024} show that MAXI~J1820+070 in the hard state outburst decay shows remarkably similar lag properties to Cyg X-1, for similar broadband noise power spectral shape. Specifically, both sources (and BHXRB candidates MAXI~J1348-630 and AT~2019wey) show a sharp `step' in the medium-soft X-ray lags at frequencies of 1--2~Hz. This timing feature has been linked to the weak and low-coherence QPO feature, which may be the analogue in the soft-hard transition of the Type C QPOs seen in transient BHXRBs in outburst rise and the hard-soft transition \citep{Alabartaetal2025,Bellavitaetal2025,Fogantinietal2025}. The power-law spectral indices of our Cyg~X-1 spectra indicate that they correspond to region Z1 (photon index $\Gamma<1.7$) of the colour-flux diagram analysed by \citet{Fogantinietal2025}, which does not yet show the emergence of this timing feature, which appears for $\Gamma>1.8$. We have checked the MAXI~J1820+070 hue values in the decaying hard state and unfortunately, due to the rapid nature of the decay, none of the available data cover the hue range of Cyg~X-1 which we study in this paper (also after correcting for mass scaling of power-colour bands used to calculate hue). Therefore at this time, we can only note the possibility that the differences in spectral-timing properties between Cyg X-1 and MAXI~J1820+070 which we report here, may be linked to their different luminosities and/or their relative locations on a hardness-intensity diagram. Cyg~X-1 may be more similar to MAXI~J1820+070 at the start of outburst decay, i.e. on the lower part of the hardness-intensity diagram (HID), than in the bright hard state (and upper track of the HID) which we study here. A more detailed comparison of Cyg~X-1 spectral-timing properties with other transient sources in outburst decay may allow a comparison of the equivalent hues, to test this possibility.

\section{Conclusions}

In this work, we have undertaken a detailed spectral-timing comparison of the black hole X-ray binaries Cyg~X-1 and MAXI~J1820+070 during their hard states, utilizing data from {\it XMM-Newton}, {\it NICER}, and {\it INTEGRAL}. Our primary aim was to compare the timing properties of these sources at two different points in their hard states, characterized by different power-spectral hues. We have revealed significant differences in the spectral-timing behaviour between the two sources at similar hues, particularly in terms of their lag-frequency and lag-energy spectra. The different black hole masses of the two systems cannot explain these differences, despite adjusting for the expected time-scale scaling with black hole mass in the frequency-dependent spectral-timing products. 

Instead, the observed discrepancies likely point to differences related to the difference in relative accretion rates and disc temperatures. However, this would require that the observed disc temperature of Cyg~X-1 is overestimated due to the presence of an additional cool Comptonizing component, which raises the question as to whether differences in the disc-corona geometry play a significant role in explaining the difference in spectral-timing properties of Cyg~X-1 and MAXI~J1820+070, even though their power-spectral shapes are broadly similar. This difference in coronal geometry may not be an intrinsic difference of Cyg~X-1 compared to transient BHXRBs, but may instead be linked to the apparent correspondence of Cyg~X-1 to the lower part of the BHXRB hardness-intensity diagram. This possibility is further supported by the recent discovery of a sharp timing feature in the lags and coherence which appears in Cyg~X-1 and multiple transient BHXRBs (including MAXI~J1820+070) in the dim (i.e. soft-hard) hard intermediate state \citep{Mendezetal2024,Koenigetal2024}.

Our work has shown the potential of comparing high-quality spectral-timing data across different black hole systems, which are selected to be in the same state and with similar power-spectral shapes. To make further progress however, it is imperative to sample and study in detail the spectral-timing properties of BHXRBs across a wide range of luminosities, to confirm whether there is a distinct but repeatable difference between spectral-timing properties of BHXRBs in the bright and dim hard states and the corresponding bright hard-soft and dim soft-hard transitions. This will require well-designed monitoring campaigns to catch the rapid hard-soft transition with sufficient exposure from X-ray timing instruments and obtain precise measurements of BHXRB spectral-timing properties at lower flux levels.

\section*{Acknowledgements}
We dedicate this work to our co-author Katja Pottschmidt who passed away while it was under review. Katja began this series of papers on Cygnus X-1 and she contributed so much to our understanding of this fascinating source. The authors would like to thank the anonymous referee for their suggestions, which improved the quality and clarity of this work. This work uses \texttt{Stingray}, a Python library developed for times-series analysis which supports a range of commonly-used Fourier analysis techniques \citep{stringray-1}, and \texttt{AstroPy}, a community-developed core Python package along with an ecosystem of tools and resources for astronomy \citep{Astropy2022}. Additionally, we have extensively utilized \texttt{NumPy} \citep{Numpy}, \texttt{SciPy} \citep{scipy}, and \texttt{Matplotlib} \citep{matplotlib} in this work. BDM acknowledges support via Ram\'on y Cajal Fellowship (RYC2018-025950-I), the Spanish MINECO grants PID2020-117252GB-I00, PID2022-136828NB-C44, and the AGAUR/Generalitat de Catalunya grant SGR-386/2021. EVL is supported by the Italian Research Center on High Performance Computing Big Data and Quantum Computing (ICSC), a project funded by European Union – NextGenerationEU – and National Recovery and Resilience Plan (NRRP) – Mission 4 Component 2 within the activities of Spoke 3 (Astrophysics and Cosmos Observations). KP was supported by NASA under award number 80GSFC24M0006. For the purpose of open access, the authors have applied a Creative Commons Attribution license to the Author Accepted Manuscript version arising from this submission.

\section*{Data Availability}

This work utilized data from the High Energy Astrophysics Science Archive Research Center (HEASARC), an online service offered by the NASA Goddard Space Flight Center, and the {\it XMM-Newton} data archive, from the European Space Agency. The data supporting this work can be accessed at HEASARC: \url{https://heasarc.gsfc.nasa.gov/docs/archive.html} and {\it XMM-Newton} archive: \url{https://nxsa.esac.esa.int/}. After the paper is published, a basic reproduction package containing the results and figures will be available on Zenodo.



\bibliographystyle{mnras}
\bibliography{mnras_template} 


\bsp	
\label{lastpage}
\end{document}